\documentclass[preprint,showpacs,superscriptaddress,preprintnumbers,amsmath,amssymb,pra]{revtex4}
\usepackage{graphicx}
\usepackage{subfigure}
\usepackage{graphicx}
\usepackage{dcolumn}
\usepackage{bm}
\usepackage{amsmath}
\usepackage{setspace}
\usepackage{mathrsfs}

\newcommand{\DvR}[1]{\left| #1 \right\rangle}
\newcommand{\loc}[0]{\mathit{l}}
\newcommand{\hmlt}[0]{\mathscr{H}}

\newcommand{\DvL}[1]{\left\langle #1 \right|}
\newcommand{\DvIP}[2]{\left\langle #1 | #2 \right\rangle}
\newcommand{\mc}[0]{J}																			  
\newcommand{\mq}[0]{\hat{J}}														    	
\newcommand{\ms}[0]{j}																				
\newcommand{\pc}[0]{\Theta}															      
\newcommand{\pq}[0]{\hat{\Theta}}															
\newcommand{\ps}[0]{\theta_m}																	
\newcommand{\sm}[1]{\left\langle(\Delta #1)^2\right\rangle}   
\newcommand{\cdc}[0]{D}																			  
\newcommand{\mi}[0]{n}      																	

\begin{document}

\title{Study of localization in the quantum sawtooth map emulated on a quantum information processor}

\author{Michael K. Henry}
\affiliation{Department of Nuclear Science and Engineering, Massachusetts Institute of Technology, Cambridge, Massachusetts 02139, USA}
\author{Joseph Emerson}
\affiliation{Perimeter Institute for Theoretical Physics, Waterloo, Ontario N2J 2W9, Canada}
\affiliation{Institute for Quantum Computing, University of Waterloo, Waterloo, Ontario N2L 3G1, Canada}
\author{Rudy Martinez}
\affiliation{Department of Natural Sciences, New Mexico Highlands University, Las Vegas, New Mexico 87701, USA}
\author{David G. Cory}
\affiliation{Department of Nuclear Science and Engineering, Massachusetts Institute of Technology, Cambridge, Massachusetts 02139, USA}

\date{\today}

\begin{abstract}
Quantum computers will be unique tools for understanding complex quantum systems.  We report an experimental implementation of a 
sensitive, quantum coherence-dependent localization phenomenon on a quantum information processor (QIP).  The localization effect was studied by emulating the dynamics of the quantum sawtooth map in the perturbative regime on a three-qubit QIP.  Our results show that the width of the probability distribution in momentum space remained essentially unchanged with successive iterations of the sawtooth map, a result that is 
consistent with localization.  The height of the peak relative to the baseline of the probability distribution did change, a result that is consistent with our QIP being an ensemble of quantum systems with a distribution of errors over the ensemble.  We further show that the previously measured distributions of control errors correctly account for the observed changes in the probability distribution.

\end{abstract}


\maketitle

The development of quantum computers promises a new approach for exploring quantum mechanics in complex systems.  In the future, we hope to use quantum computation to emulate quantum behavior in Hilbert spaces that are larger than can be simulated on a classical computer.  Today we have access to QIPs that are prototypes of quantum computers (QCs).  These devices operate over small and limited Hilbert spaces, and with significant noise.  However, even with these limitations they can be used to explore questions of quantum mechanics that start to reflect the power we expect of future QCs.  Even when QIPs operate on highly mixed states, as is the case for liquid state NMR implementations, we can distill properties that are consistent with the desired quantum phenomena.  Here we will explore one such example, localization under the quantum sawtooth map.  Localization is a uniquely quantum phenomenon and thus a natural target of quantum computation.  

Dynamical localization occurs in classically chaotic quantum maps, in which the quantum state initially diffuses at the classical rate due to repeated quasi-random perturbations, but then stabilizes to a fixed probability distribution and remains coherently localized under subsequent perturbations \cite{Casati1979}.  In the case of perturbative localization, the probability distribution is localized with no initial period of diffusion.  In both cases, the exponentially peaked, static probability distribution distinctly contrasts with the classical, diffusive behavior.  Efficient algorithms have been developed for simulating the dynamics of localization in the kicked rotator model \cite{Georgeot2001}, the quantum sawtooth map \cite{Benenti2001}, and the kicked Harper model \cite{Levi2004} on a QIP.  Although localization can in principle be observed in an ideal emulation using as few as three qubits \cite{Benenti2004}, the sensitivity of this phenomenon to noise effects \cite{Ott1984, Benenti2001, Benenti2004, Lee2005, Song2001, Levi2003, Levi2004} poses a rigorous challenge in the task of creating and maintaining a localized state on a noisy QIP.  

In our experiment, we explored localization on a small (3-qubit) QIP based on liquid state NMR.  The aim of the study was to implement the sawtooth map on our QIP, to do so in such a way that we observe properties of localization which are clearly distinct from the classical behavior, and to use this example emulation as a test of the precision of our implementation as well as to motivate the continued refinement of this implementation. 

Since liquid state NMR QIP relies on a large spatially distributed ensemble of quantum systems, we have to be careful in selecting the measures that we use for observing localization.  The errors in our implementation of the sawtooth map will vary over the ensemble due to the inhomogeneity of the rf control field.  There will be regions of the ensemble where the fidelity of implementation of the sawtooth map is sufficient that we observe localization, while for other regions the errors will be large enough to prevent localization.  In the experiment we observe the sum of these effects, and thus we expect to see a peak in the probability distribution in the basis of localization that is representative of those parts of the ensemble that are localized, accompanied by a background offset in the probability distribution in the basis of localization resulting from those parts of the ensemble that are not localized. 

A description of the sawtooth map is given in Sec. \ref{sec:sawtooth}, followed by an explanation of the experimental implementation of the map in Sec. \ref{sec:imp_det}.  A discussion of the noise effects expected in the experiment are presented in Sec. \ref{sec:num_sim}, along with a discussion of their effects on the localization phenomenon.   In Sec. \ref{sec:results} the experimental results are reported and shown to demonstrate properties which are consistent with localization.  The results are further compared with numerical simulations of the experiment which show that the imperfections in the data are well accounted for by the error model identified in previous work \cite{Weinstein2004, Pravia2003, Boulant2004}.  Finally, in Sec. \ref{sec:discussion}, numerical studies of the error model are applied to measure the relative degree of delocalization caused by the specific noise mechanisms. 

\section{The sawtooth map}
\label{sec:sawtooth}

The sawtooth map is a periodically kicked system with period $T$ and kick strength $k$, whose classical dynamics are dictated by a single parameter $K=kT$.  One iteration of the classical sawtooth map is compactly described by the equations
\begin{align}
\label{eq:cl_map}
\bar{\mc}&=\mc+k(\pc-\pi) \\ \bar{\pc}&=\pc+T\bar{\mc} \nonumber
\end{align}
where $\pc$ is the angular position variable and $\mc$ is the angular momentum variable.  The cylindrical phase space, which results from the periodicity of the position variable ($0\leq\pc<2\pi$), can be represented on a torus by truncating the momentum space to length $2\pi L/T$ and applying a periodic boundary condition.  In the quantum regime, one iteration of the sawtooth map is represented by the unitary time evolution operator
\begin{equation}
U_{saw}(0,T)=\exp\left(-iT\mq^2/2\right)\exp\left(ik(\pq-\pi)^2/2\right)
\end{equation}
where $\mq$ and $\pq$ are conjugate action quantum mechanical operators.  The state of the quantum system is represented by a density matrix $\hat{\rho}$ expressed in the momentum basis.  A detailed description and insightful discussion of the sawtooth map can be found in reference \cite{Benenti2004}.  

In a simulation of the quantum sawtooth map on an $n_q$ qubit quantum information processor, the momentum basis states of the emulated system are represented by $N=2^{n_q}$ computational basis states, therefore $N=2\pi L/T$.  The momentum basis states are labeled by their eigenvalues $-N/2\leq \ms < N/2$, such that $\mq\DvR{\ms}=\ms\DvR{\ms}$.  The position basis states $\DvR{\ps}$ have eigenvalues $\ps=(2\pi m/N)$, such that $\pq\DvR{\ps}=\ps\DvR{\ps}$, where $0\leq m<N$.  The overlap between conjugate basis states is given by
\begin{equation}
\DvIP{\ps}{\ms}=\frac{1}{\sqrt{N}}\exp\left[\frac{2\pi im}{N}\left(\ms+\frac{N}{2}\right)\right]  
\end{equation}

\subsection{Localization in the quantum sawtooth map}
\label{sec:loc_sawtooth}

\begin{figure}
	\begin{center}
 	\includegraphics[width=3.2in]{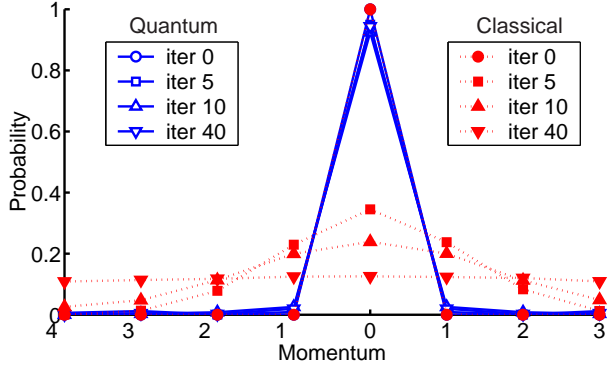}
 	\end{center}
	\caption{\label{classical_v_quantum}  The momentum distribution after 0, 5, 10, and 40 iterations of the classical (red, filled markers) and quantum (blue, unfilled markers) sawtooth map ($L=7$, $K=1.5$, $N=8$).  The classical distribution represents 20,000 realizations of the map, with initial momenta $\ms=0$ and random initial positions uniformly distributed over the phase space.  The initial quantum state is the $\ms=0$ momentum eigenstate.  In the quantum case, the momentum is discrete, and each data point represents the population of the indicated momentum state.  In the classical case, the momentum is continuous, and each data point represents the probability of momentum being in the range of the indicated value $\pm 1/2$. The classical map is chaotic, which leads to the observed diffusive behavior.  In the quantum map, since the localization length is less than 1, the state remains exponentially localized after a single iteration.  The breadth of the momentum distribution is essentially static in the quantum case, and only the $\ms=0$ momentum state is significantly populated.}
\end{figure}

In the classical phase space, when $K<-4$ or $K>0$, the sawtooth map induces chaotic motion, which is seen by considering a classical ensemble of trajectories, where each element of the ensemble has a fixed initial momentum ($\mc=0$) and a randomized initial position ($\pc$).  The chaotic motion arises due to the presence of the term $k \left(\pc - \pi\right)$ in Eq. \ref{eq:cl_map}, which gives a \textit{kick} to the momentum at each iteration of the map.  In chaotic parameter regimes, the strength of the sequence of kicks can be approximated as a quasi-random sequence, leading to diffusive broadening along the momentum dimension of the classical phase space, as shown in Fig. \ref{classical_v_quantum}.  As a result, the breadth of the distribution, as measured by its second moment, grows linearly with the number of map iterations, $\mi$, according to
\begin{equation}
\sm{\mc}\approx \cdc \mi,
\end{equation}
where $\cdc \approx (\pi^2/3)k^2$ is the classical diffusion coefficient.  As the map is iterated, momentum diffusion continues indefinitely, and the probability distribution approaches uniformity over the bounded toroidal phase space.
   
The quantum system demonstrates a strikingly different behavior.  Like the classical map, the quantum sawtooth map initially causes diffusive broadening in the momentum basis according to the classical diffusion coefficient $\cdc$.  However, after $\mi^*\approx \cdc$ iterations of the quantum map, diffusion is suppressed due to quantum interference, and for all subsequent iterations, the quantum state maintains roughly the same exponentially localized profile over the momentum basis.  This surprising interference effect requires the coherence of the quantum state.  The square root of the quantum standard deviation $\sqrt{\sm{\mq}}$ represents the number of states that are significantly populated in the system, and is essentially static after $\mi^*$ iterations, representing the onset of localization.  Therefore, the localization length
\begin{equation}
\loc=\sqrt{\sm{\mq}}=\sqrt{\cdc\mi^*} 
\label{eq:loc_def}
\end{equation}
serves as a useful parameter for characterizing a localized state.

The heuristic approximation $\mi^*\approx \cdc \approx \loc$ \cite{Benenti2003} yields a theoretical prediction of the localization length based on the kick strength $\loc\approx(\pi^2/3)k^2$.  Quantum localization occurs when the localization length is less than the total breadth of the phase space $N$.  However, the degree to which a physical system becomes localized may be reduced by noise effects, as this unique quantum phenomenon is quite sensitive to decoherence and other types of errors \cite{Ott1984, Song2001, Levi2003, Lee2005}.   This sensitivity poses a rigorous challenge when trying to create and maintain a localized state on a noisy QIP. 

The implementation of the quantum sawtooth map reported here takes $L=7$, $K=1.5$, $N=8$, which corresponds to the classically chaotic regime, with a diffusion coefficient of $\cdc\approx(\pi^2/3)k^2=0.24$.  The theoretical approximation that $\mi^*\approx\cdc$ predicts that the system will be localized after one iteration of the map ($\mi^*<1$).  This effect is known as perturbative localization, where the system is localized without the initial diffusive behavior.  The results of numerical simulations plotted in Fig. \ref{classical_v_quantum} confirm this prediction, as the breadth of the probability distribution is essentially static after a single iteration of the quantum map.  

\section{Implementation details}
\label{sec:imp_det}

An algorithm for the quantum sawtooth map can be generated by expressing the matrix elements of the map in the momentum basis:
\begin{widetext}
\begin{align}
\DvL{\ms}U_{saw}\DvR{\ms^{\prime}} 
		&=\sum_{m}\DvL{\ms}\exp\left(-iT\mq^2/2\right)\DvR{\ps}
		\DvL{\ps}\exp\left(ik(\pq-\pi)^2/2\right)\DvR{\ms^{\prime}} \nonumber \\
	 	&=\sum_{m}\exp\left(-iT\ms^2/2\right)\DvIP{\ms}{\ps}
	 	\exp\left(ik(\ps-\pi)^2/2\right)\DvIP{\ps}{\ms^{\prime}} \nonumber \\
	 	&=\DvL{\mc}U_{\mc}U_{QFT}^{-1}U_{\pc}U_{QFT}\DvR{\ms^{\prime}}
\end{align}
\end{widetext}
where $U_{QFT}$ is the familiar Quantum Fourier Transform (QFT) which has the action of toggling between the position and momentum basis representations, and the diagonal free evolution and kick operators, $U_{\mc}$ and $U_{\pc}$, are defined
\begin{align}
\DvL{\ms}U_{\mc}\DvR{\ms^{\prime}} & =\exp\left(-iT\ms^2/2\right)\DvIP{\ms}{\ms^{\prime}} \\
\DvL{\ps}U_{\pc}\DvR{\ps^{\prime}} & =\exp\left(ik(\ps-\pi)^2/2\right)\DvIP{\ps}{\ps^{\prime}}.
\end{align}

This form of the quantum sawtooth map ($U_{saw}=U_{\mc}U_{QFT}^{-1}U_{\pc}U_{QFT}$) reveals the underlying structure of the map:  After the system is initialized to a momentum basis state, the first operation in the quantum sawtooth map, the QFT, transforms the system to the position basis representation, where the diagonal kick operator $U_{\pc}$ applies an impulse force.  The inverse of the QFT is applied next, returning the system to the momentum basis representation, where the diagonal free evolution operator $U_{\mc}$ is applied.  These four steps ($U_{QFT}, U_{\pc}, U_{QFT}^{-1}, U_{\mc}$) constitute a single iteration of the quantum sawtooth map.  After iterating the map, the localized probability distribution corresponds to the diagonal elements of the density matrix, $W_{\ms}\equiv\DvL{\ms}\hat{\rho}\DvR{\ms}$.  Realizing that the only effect of the free evolution operator is to apply a phase to the coefficient of each momentum basis state, $U_{\mc}$ can be neglected in the final iteration of the map, since a phase does not alter the measured probabilities in that basis.

The quantum circuit for the QFT is given in \cite{MikeIke}, an implementation of the QFT is described in \cite{Weinstein2001}, and an analysis of the quantum process tomography of this implementation can be found in \cite{Weinstein2004}.  Circuits for $U_{\mc}$ and $U_{\pc}$ are conveniently found in realizing that the diagonal operators can be decomposed into a series of single-qubit z rotations and two-qubit controlled z rotations; the details of this decomposition are given in \cite{Benenti2004}.  This circuit for the quantum sawtooth map is computationally efficient, in that the number of fundamental quantum gates required to implement the algorithm depends polynomially on the number of qubits \cite{Benenti2001}.  

\subsection{Quantum information processing with NMR}
\label{sec:QIP_NMR}

\begin{figure}
	\begin{center}
 	\includegraphics[width=3.2in]{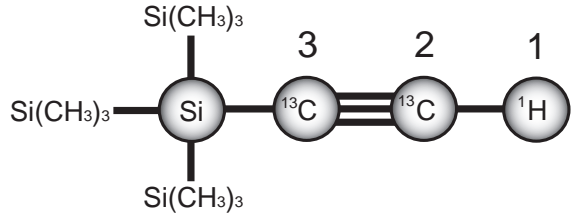}
 	\end{center}
	\caption{\label{TMSA} A diagram of the tris(trimethylsilyl)silane-acetylene molecule used to simulate the quantum sawtooth map in a liquid state NMR QIP.  The hydrogen nucleus in the acetylene branch is labeled qubit 1 in the experiment; the two carbon-13 nuclei in the acetylene branch are labeled qubits 2 and 3.}
\end{figure}

In NMR quantum information processing, nuclear spins polarized by a strong external magnetic field serve as qubits.  The molecule used in this experiment, diagrammed in Fig. \ref{TMSA}, is tris(trimethylsilyl)silane-acetylene dissolved in deuterated acetone.
The carbon nuclei in the acetylene branch are carbon-13 enriched, and the methyl carbons are of natural isotopic abundance.  The two carbon-13 nuclei and the hydrogen nucleus in the acetylene branch are used as qubits.  The full internal Hamiltonian has the form
\begin{equation}
\label{HintFull}
\hmlt_{int}=\sum_{i=1}^{n_q}{\omega_i\sigma_z^i}+\sum_{j<k}{J_{jk}\sigma^j\cdot\sigma^k}
\end{equation}
where $\omega_i$ is the resonance frequency of the $i^{th}$ spin, and $J_{jk}$ is the frequency of scalar coupling between spins $j$ and $k$.  The hydrogen nucleus is labeled qubit number $1$, making it the most significant bit in the computational state vector.  The carbon qubits are labeled as indicated in Fig. \ref{TMSA}.  Experiments are performed in a 9.4 Tesla magnetic field, where the Carbon qubits are separated by 1.201 kHz.  The scalar couplings are $J_{12}=235.7$ Hz, $J_{23}=132.6$ Hz, and $J_{13}=42.9$ Hz.  Because the spin system is in a highly mixed state at room temperature, the system was prepared in a pseudopure state \cite{Cory1997} by the technique described in \cite{Teklemariam2001}.  Three readout sequences were needed to measure the eight diagonal elements of the density matrix, which correspond to the distribution of momentum basis states.

\subsection{Implementing the quantum sawtooth map}
\label{sec:imp_qsm}

\begin{figure*}
	\begin{center}
	\includegraphics[width=6.4in]{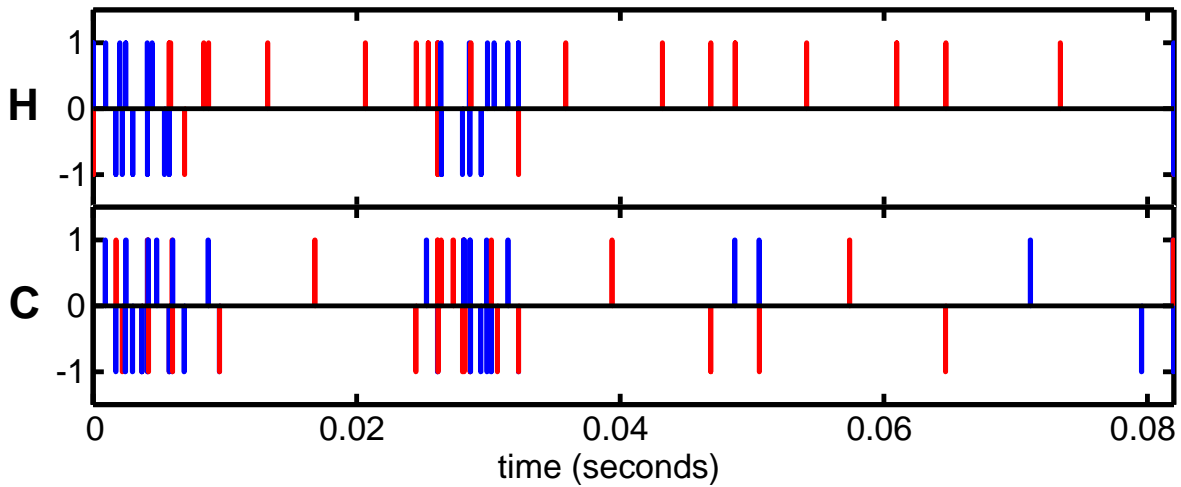}
	\end{center}
	\caption{The hydrogen (H) and carbon (C) rf control fields versus time for the full quantum sawtooth map pulse sequence.  Red versus blue pulses are 90 degrees out of phase; pulses above versus below the horizontal axis are 180 degrees out of phase.  The nutation frequency of each hydrogen (carbon) pulse is 293.6 kHz (108.9 kHz).}
	\label{fig:pulse_seq}
\end{figure*}

The average gate fidelity \cite{Fortunato2002} of each unitary control sequence was optimized over the full Hilbert space.  The input state preparation pulse sequence, which is non-unitary, was optimized based on the state correlation \cite{Fortunato2002} between the simulated input state $\hat{\rho}_{sim}$ and the ideal input state $\hat{\rho}_{ideal}=\DvR{100}\DvL{100}$.  The average fidelity (or state correlation) of each implemented pulse sequence, as calculated by numerical simulation, is listed in Table \ref{fidels}.
\begin{table}
\caption{\label{fidels} Pulse sequences designed for the quantum sawtooth map experiment.  Fidelities are calculated by numerical simulations which account for rf inhomogeneity, neglecting decoherence effects.}
\begin{center}
\begin{tabular}{c|c|c}
Map & Duration (ms) & Fidelity \\
\hline\hline
Input State Preparation & 50 &  $Corr=0.99$\\
QFT & 6 &  0.99\\
QFT Inverse & 6  & 0.99\\
$\mc$ Diagonal & 50 & 0.99\\
$\pc$ Diagonal & 20 & 0.99\\
Readout 1 & 0.01  & 1.00  \\
Readout 2 & 62  & 0.98  \\
Readout 3 & 72  & 0.98  \\
\end{tabular}
\end{center}
\end{table}
The NMR pulse sequence for one full iteration of the quantum sawtooth map is shown in Fig. \ref{fig:pulse_seq}.  

\section{Numerical Simulation}
\label{sec:num_sim}

Through numerical simulation of the experiment, it is possible to predict the behavior of the system under the optimized control sequence in the presence of various types of noise known to influence the QIP.  Errors affecting the implementation of the quantum sawtooth map are conveniently classified in three categories \cite{Pravia2003, Weinstein2004, Boulant2004} -  coherent errors, decoherent errors, and incoherent errors - which can be generally used to categorize the errors affecting any QIP.  Each type of error delocalizes the system in a different manner.   

In the presence of coherent errors, the system evolves under a unitary process other than the ideal quantum sawtooth map.  Due to their unitary nature, coherent errors are reversible.  The coherent errors modeled in numerical simulations arise in the experiment due to strong coupling between the carbon qubits, as well as the action of the internal Hamiltonian during rf pulses.  Coherent errors delocalize the system by introducing unitary transitions between momentum states.  

Decoherent errors cause the individual members of the ensemble (and hence the observed ensemble average) to evolve in a non-unitary fashion.  Decoherent evolution can be modeled as a coupling between the system and an external environment and can usually be represented by a completely positive linear map, expressed as an $N^2 \times N^2$ superoperator $S$ acting on a columnized $N^2 \times 1$ state vector $\DvR{\rho}$, according to 
\begin{equation}
\label{eq:supopevol}
\DvR{\rho_{out}}=S\DvR{\rho_{in}}.
\end{equation}
Decoherent errors are accounted for in numerical simulations by allowing the system to evolve under an approximate relaxation superoperator \cite{ErnstText}, which is completely diagonal in the generalized Pauli basis.  In this diagonal form, each non-zero entry in the relaxation superoperator represents the decoherence rate of a generalized Pauli basis operator; the specific values used in simulations are based on measurements of all $T_1s$ in the three qubit system as well as the single species $T_2s$.  

Given the time scale of one full iteration of the quantum sawtooth map ($10^{-1}$s) compared to the system's typical decoherence rates ($1$ s$^{-1}$), decoherence in this system is in the moderately dissipative regime, which has been shown to cause delocalization in the quantum sawtooth map \cite{Lee2005}.  The non-unitary, dissipative action of decoherence along with the mixing action of the control sequence causes an essentially uniform damping of the measured probability distribution, accompanied by a uniform background offset that conserves probability.  The background offset in probability appears in the density matrix as an increased identity component, which represents a loss of system purity and a corresponding increase in the von Neumann entropy of the system.  In this way, the effect of decoherence is to mimic the diffusive, chaotic dynamics of the classical system described in Sec. \ref{sec:sawtooth}. 

\begin{figure}
	\begin{center}
  \includegraphics[width=3.2in]{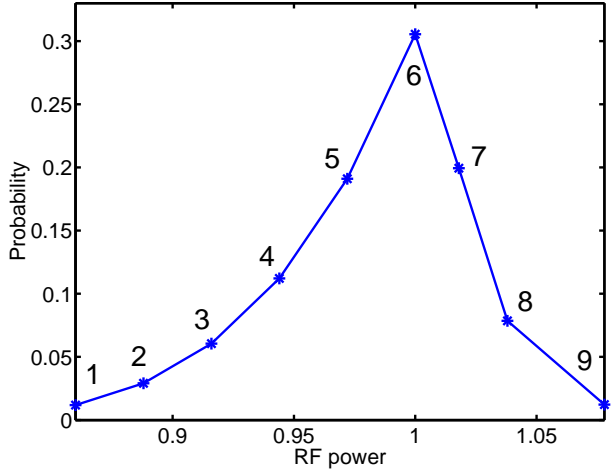}
	\end{center}
	\caption{The carbon rf probability distribution measured in previous experiments and used to generate the plots in Fig. \ref{fig:bins}.  The carbon rf power is in units of the nominal carbon nutation frequency, 108.9 kHz.  The numerals labeling each point in the distribution indicate the associated bin of incoherence in Fig \ref{fig:bins}.}
	\label{fig:rf_pdf}
\end{figure}

Incoherent errors occur when the various members of the experimental ensemble experience a distribution of unitary time evolution operators.  Incoherent evolution can be generally expressed as an operator sum
\begin{equation}
\rho_{out}^{inc}=\sum_{k}p(k)U_k\rho_{in}U_k^{\dagger},
\end{equation} 
where $p(k)$ is the probability that a member of the ensemble will undergo unitary time evolution under $U_k$.  Under incoherent errors, the individual members of the ensemble evolve coherently, but the ensemble-averaged time evolution of the system is non-unitary.  The dominant incoherent errors in the experiment arise due to the inhomogeneity of the rf field over the spatial extent of the liquid state NMR sample.  When an rf pulse is applied during the experiment, the members of the ensemble experience a distribution of rf powers, and only a fraction of the ensemble actually experiences precisely the nominal (ideal) rf power.  In numerical simulations, we can approximate the effects of the continuous distribution of carbon rf powers by simulating a previously measured distribution of nine discrete bins of rf power, plotted in Fig. \ref{fig:rf_pdf}.  
Bin 6 represents the largest portion of the ensemble and corresponds to the nominal rf power, while the other eight bins result from the inhomogeneity of the rf control field.  In the experiment, there are two dimensions of incoherence: one for both the proton and carbon rf-control fields.  In numerical simulations presented in Sections \ref{sec:results} and \ref{sec:discussion}, the continuous distribution of carbon and hydrogen rf power correlations is approximated by a discrete two-dimensional ($9 \times 9$) rf probability distribution function.  

\begin{figure}
	\begin{center}
	\begin{tabular}{ccc}
  \subfigure[Bin 1]{\includegraphics[width=2.0in]{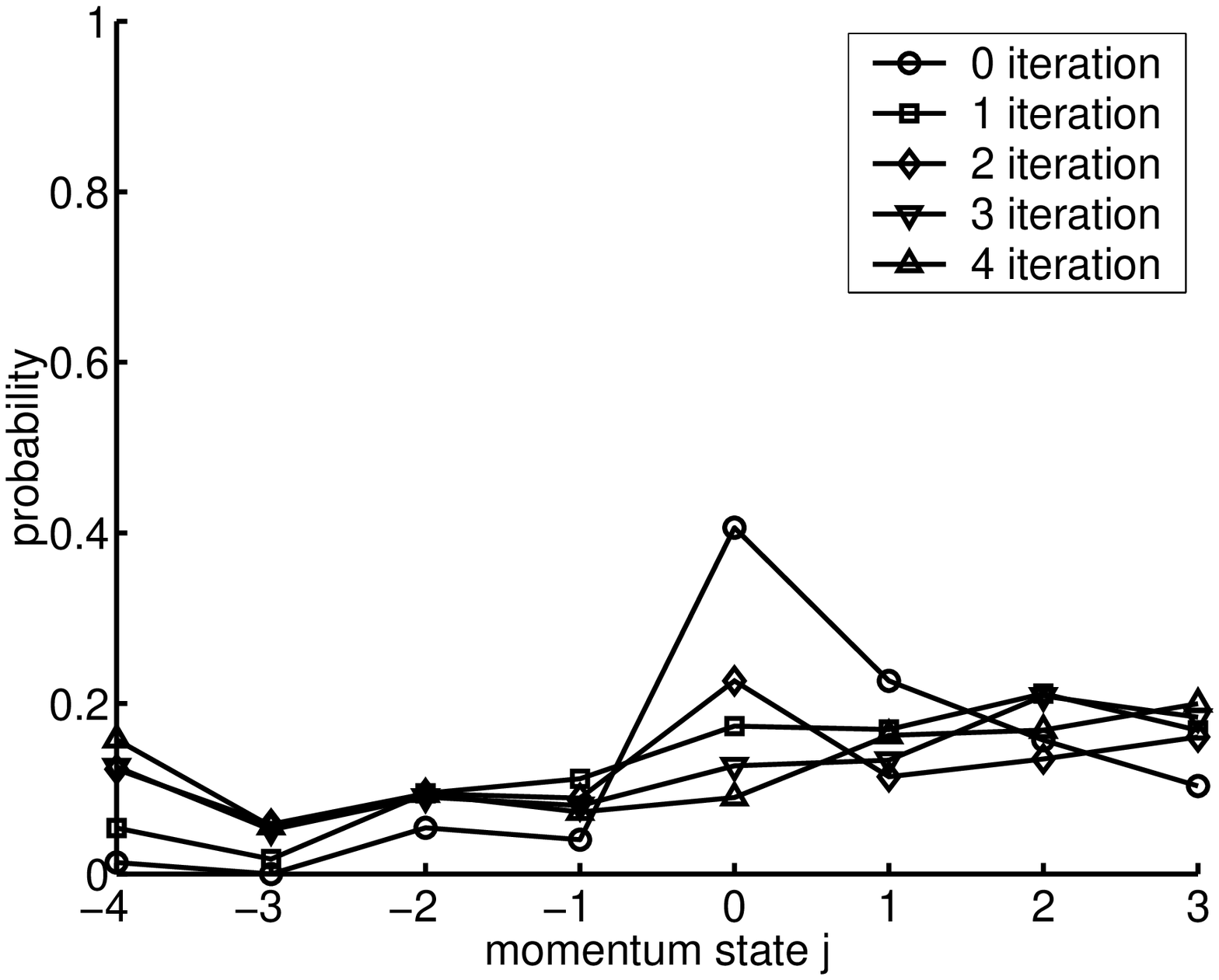}} &
  \subfigure[Bin 2]{\includegraphics[width=2.0in]{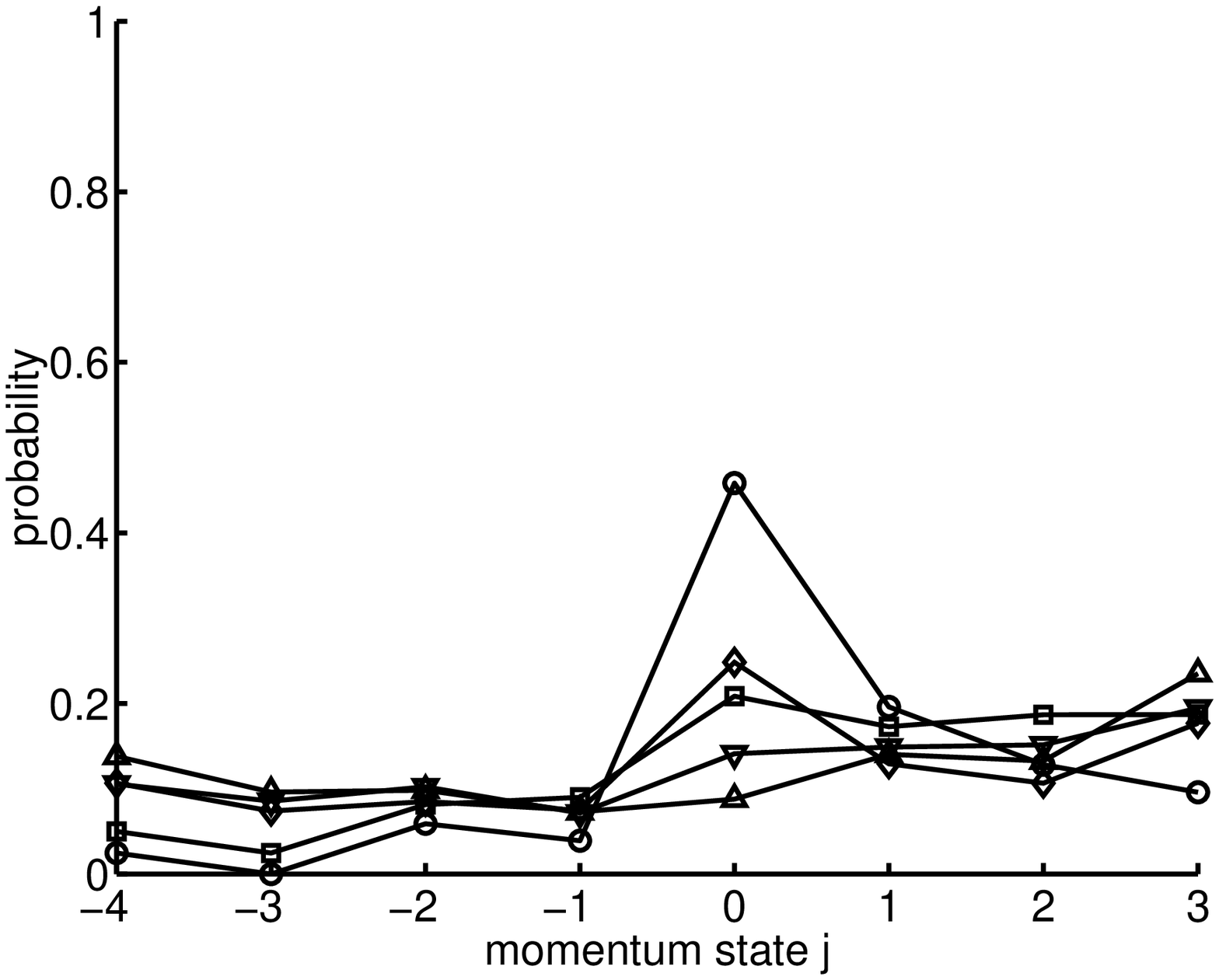}} &
  \subfigure[Bin 3]{\includegraphics[width=2.0in]{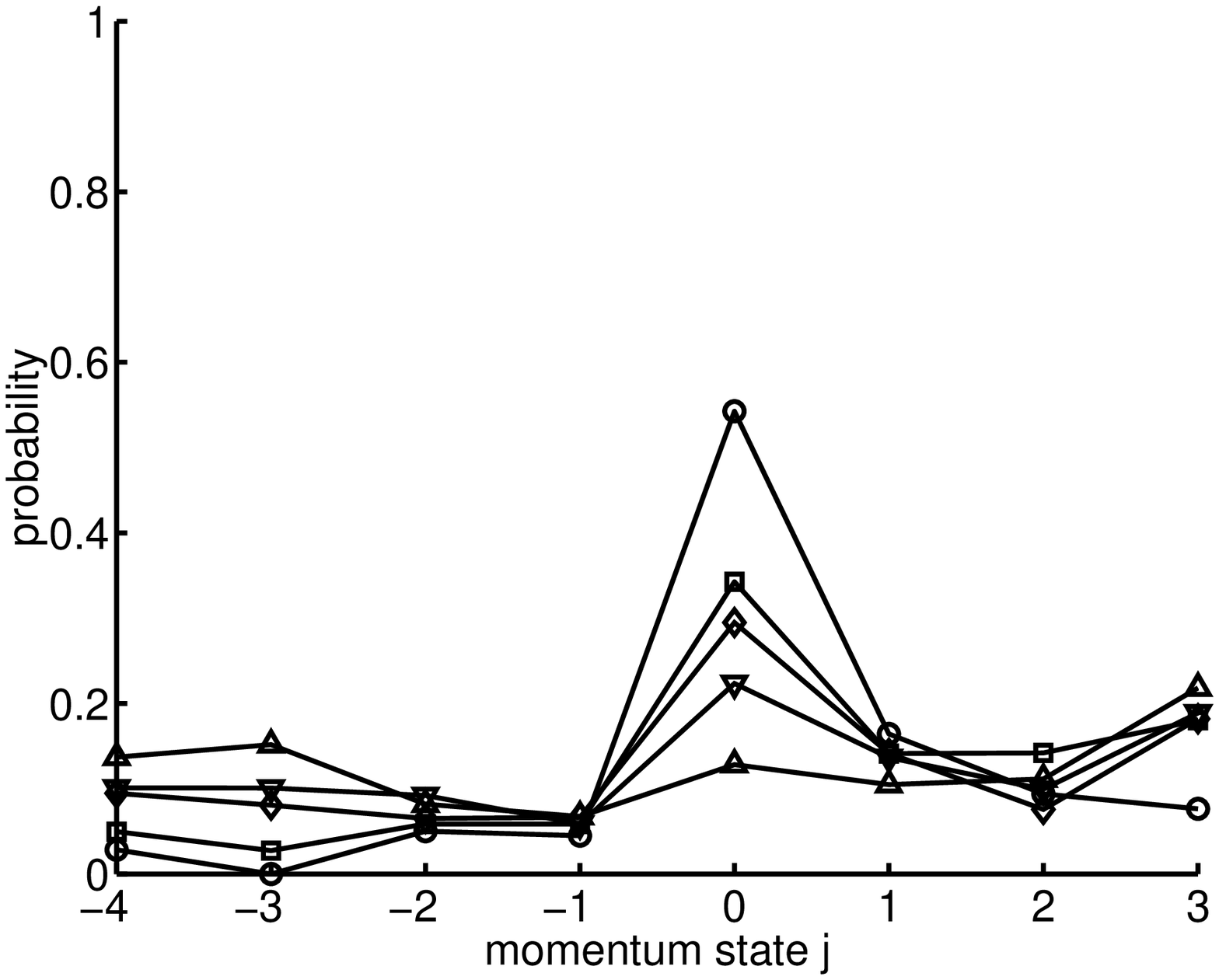}} \\
  \subfigure[Bin 4]{\includegraphics[width=2.0in]{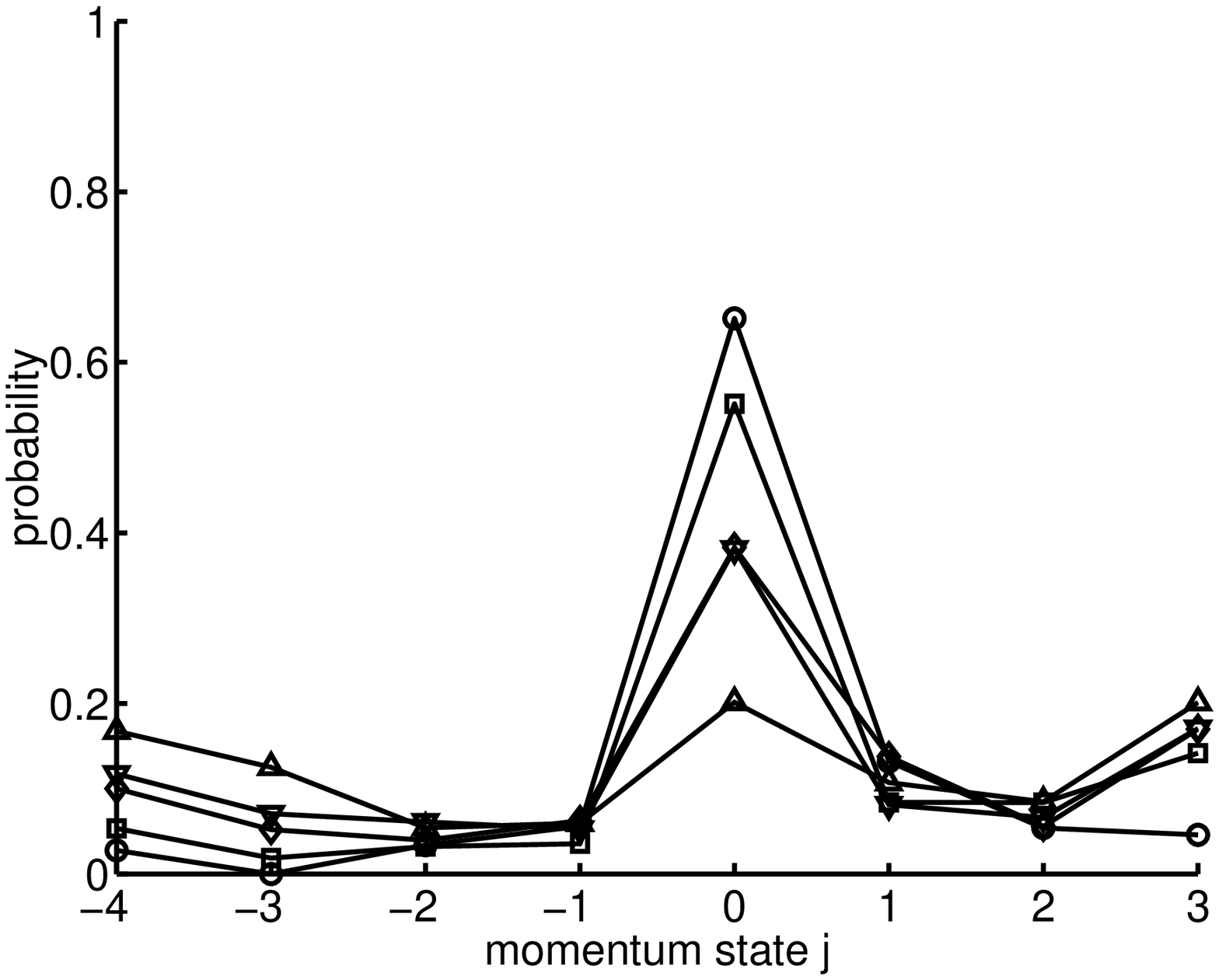}} &
  \subfigure[Bin 5]{\includegraphics[width=2.0in]{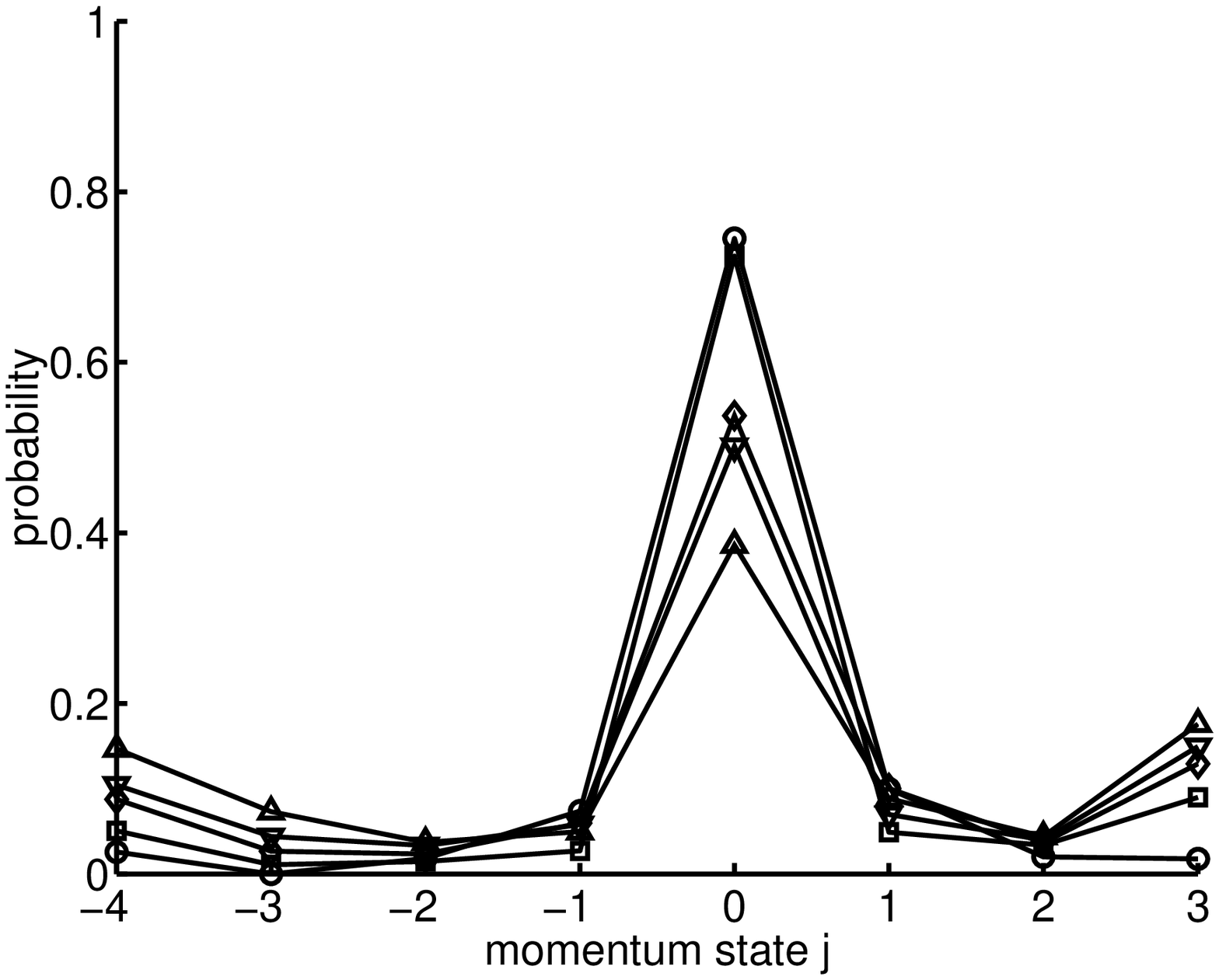}} &
  \subfigure[Bin 6]{\includegraphics[width=2.0in]{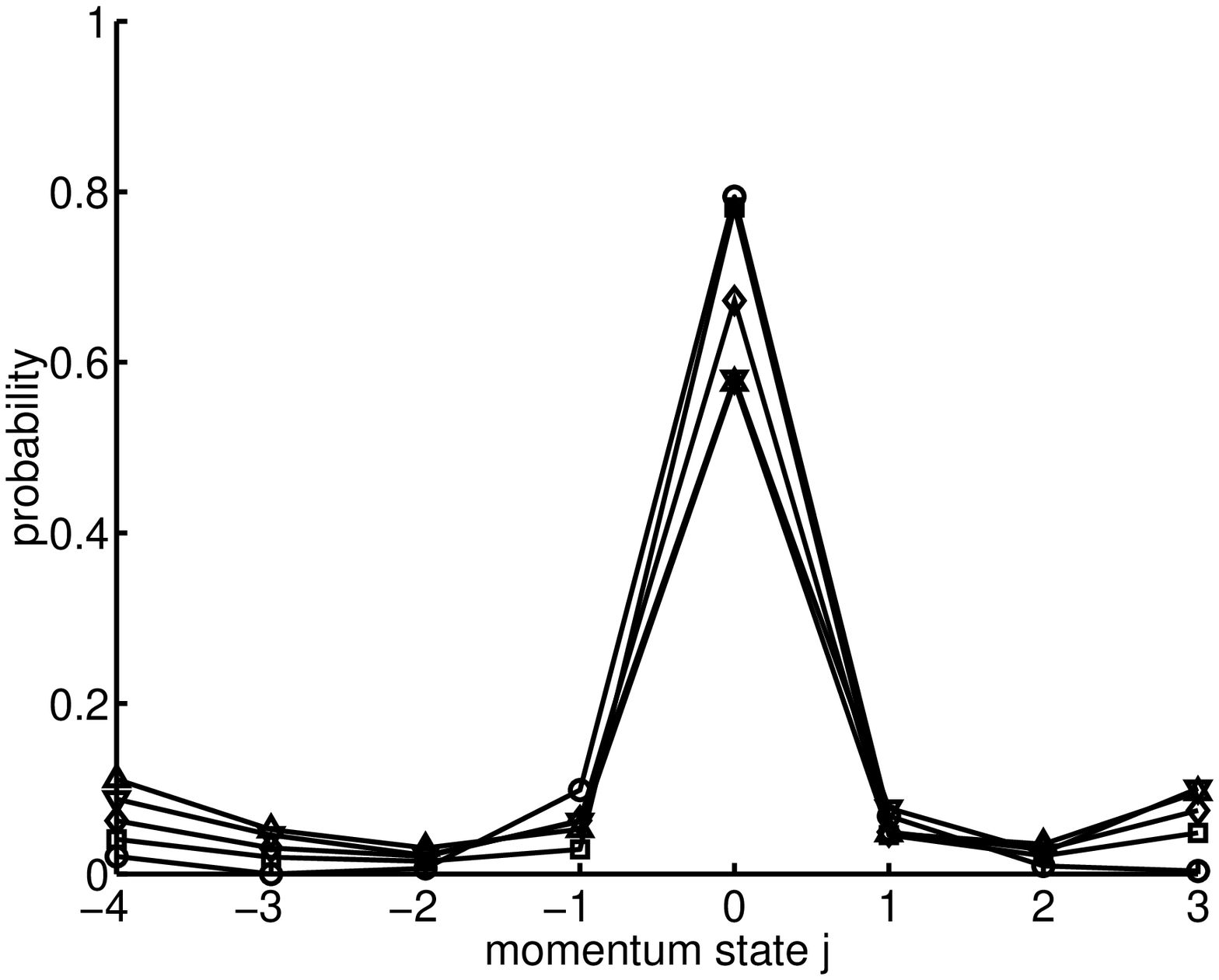}} \\
  \subfigure[Bin 7]{\includegraphics[width=2.0in]{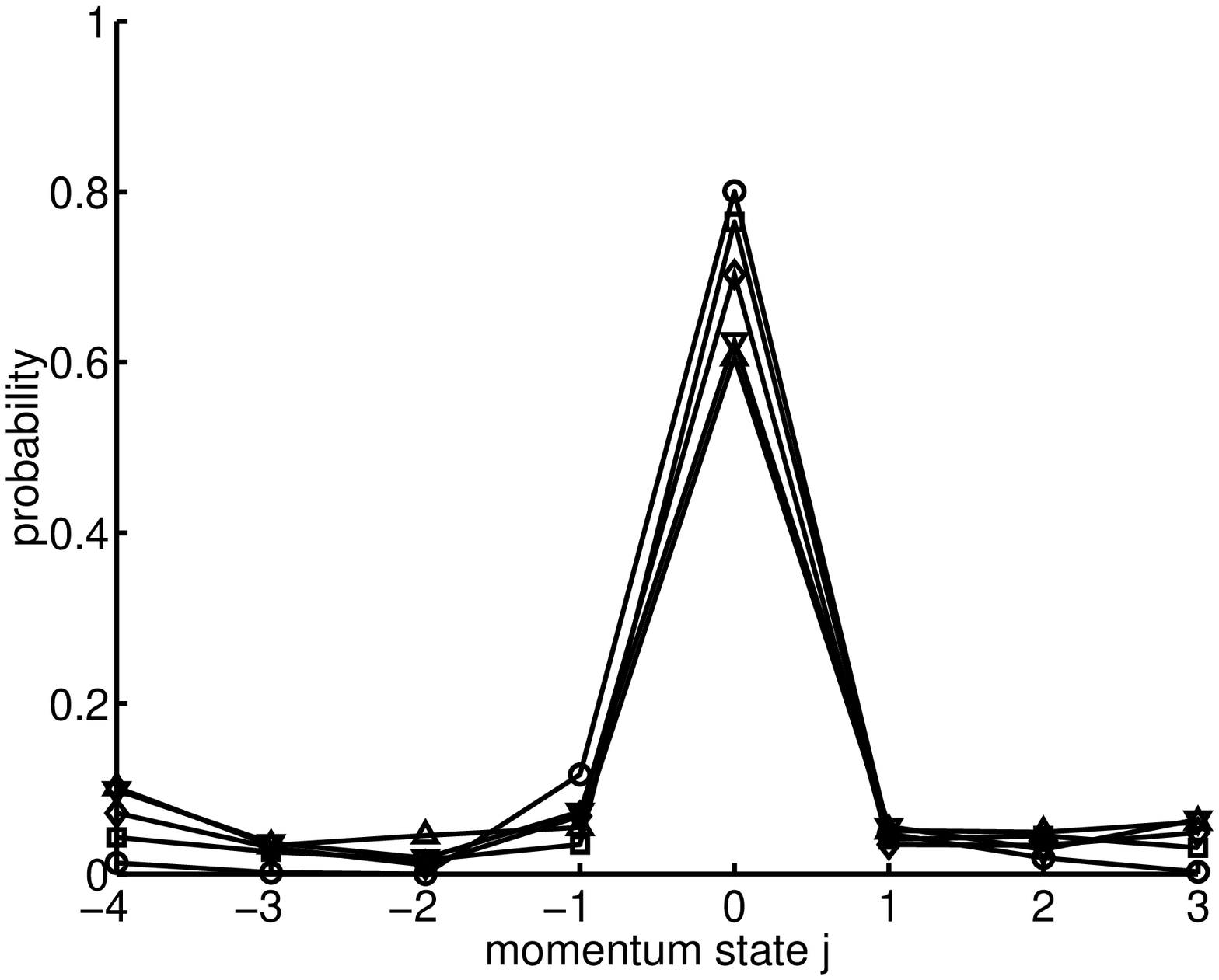}} &
  \subfigure[Bin 8]{\includegraphics[width=2.0in]{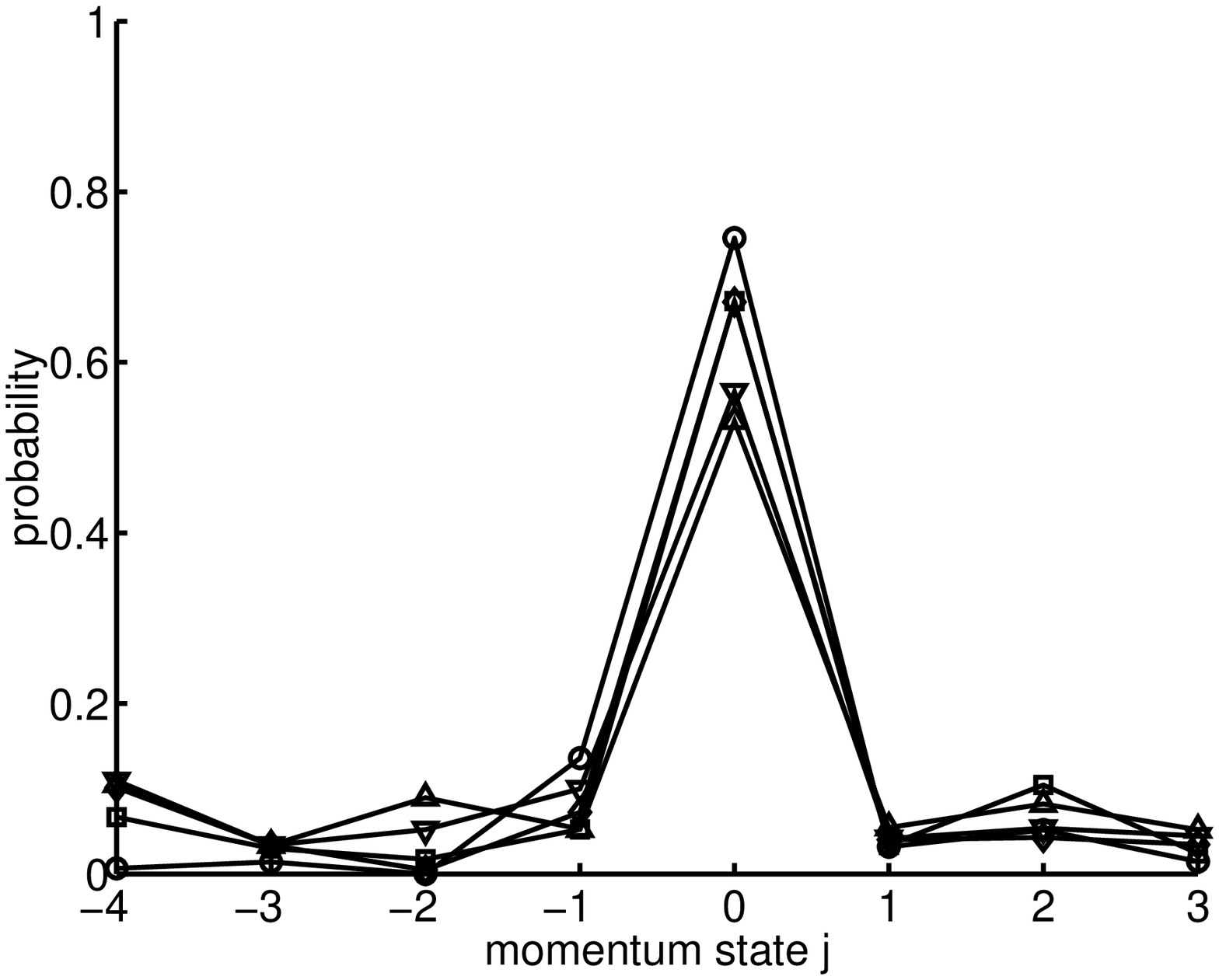}} &
  \subfigure[Bin 9]{\includegraphics[width=2.0in]{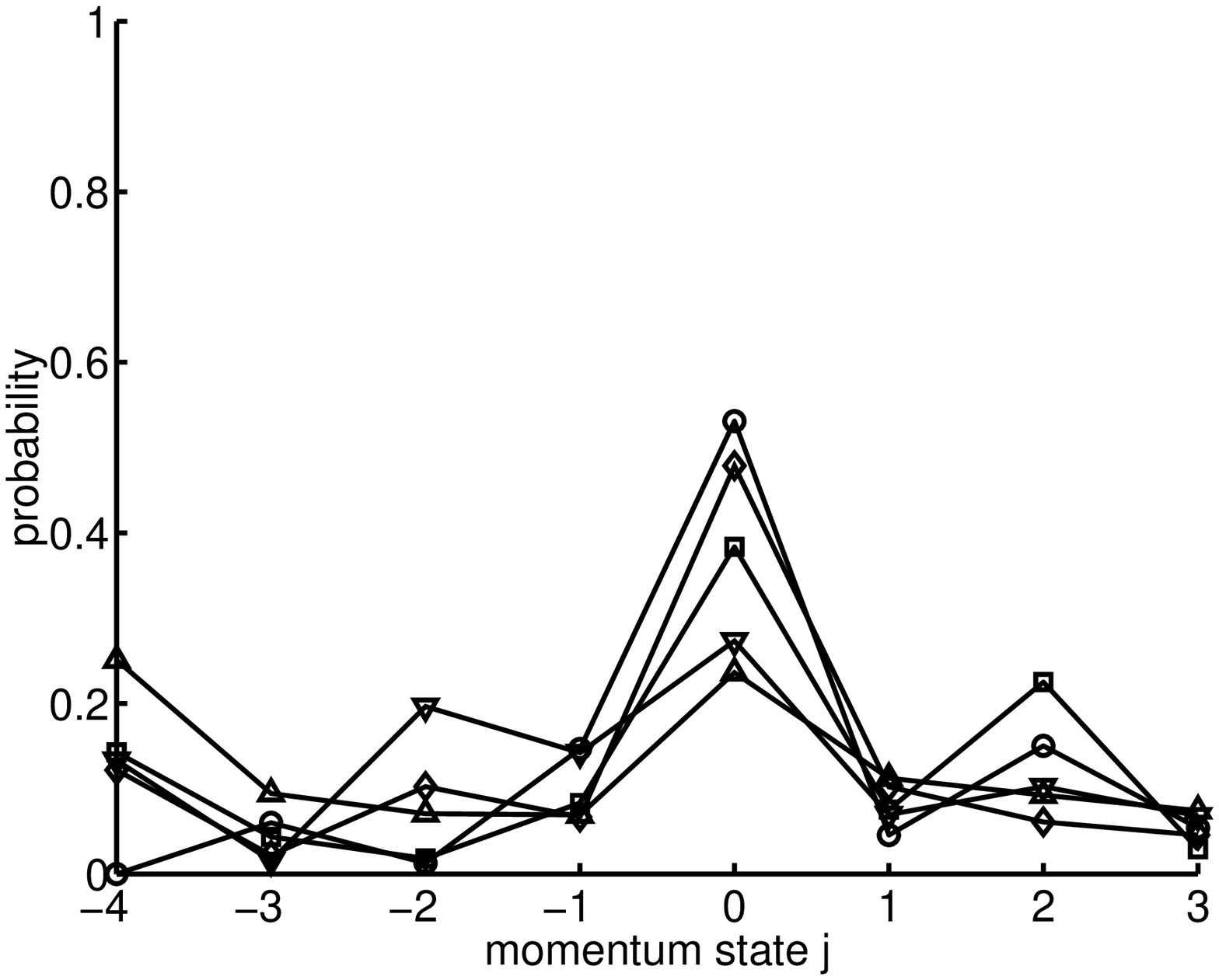}} 
  \end{tabular}
	\end{center}
	\caption{Momentum profiles in different regions of the ensemble generated by numerical simulations which also account for decoherence effects.  Each plot represents the momentum profile resulting from a numerical simulation of the control sequence (zero through four iterations) with the carbon rf power indicated in Fig. \ref{fig:rf_pdf}.  Bin 6 represents the nominal rf power, which has the highest fidelity when compared to the ideal quantum sawtooth map.  The distributions simulated near the nominal rf power appear to be localized, while those far from the nominal rf power are quickly delocalized.  In experiments, the average over the incoherence is observed (weighted by the probability distribution in Fig. \ref{fig:rf_pdf}).}
	\label{fig:bins}
\end{figure}

Simulating only the one-dimensional distribution of carbon rf power in Fig. \ref{fig:rf_pdf} is sufficient for gaining a qualitative understanding of the effects of incoherent noise in the experiment.  Figure \ref{fig:bins} 
shows the results of numerical simulations of the experimentally implemented control sequence for each bin of rf power; the simulations include decoherence effects.  Due to the incoherence, the local errors are different for each bin.  Consequently, in regions of the ensemble where the rf power is near the nominal rf power (see Bins 6 through 8), the fidelity of implementation of the sawtooth map is sufficient that we observe localization.  For other regions which constitute a smaller percentage of the ensemble (e.g. Bin 1), the errors are large enough to prevent localization and the momentum distribution is broad.  Hence we see that in the experiment, when we observe the weighted average of these distributions, the (more abundant) localized portions of the ensemble will appear as a peak in the $\ms=0$ momentum state, and the delocalized portions of the ensemble will contribute an approximately uniform background offset across the momentum distribution.  

Another important insight gained from analyzing distributions plotted in Fig. \ref{fig:bins} is that the bins where the rf power is near ideal are relatively unaffected by decoherence.  Consequently, we expect incoherence rather than decoherence to be the principle source of noise compromising our ability to observe localization over the ensemble.  This is discussed in more depth in Sec. \ref{sec:discussion}.

\section{Results}
\label{sec:results}

\begin{figure}
	\begin{center}
	\begin{tabular}{cc}
	\subfigure[input state]{\includegraphics[width=3.0in]{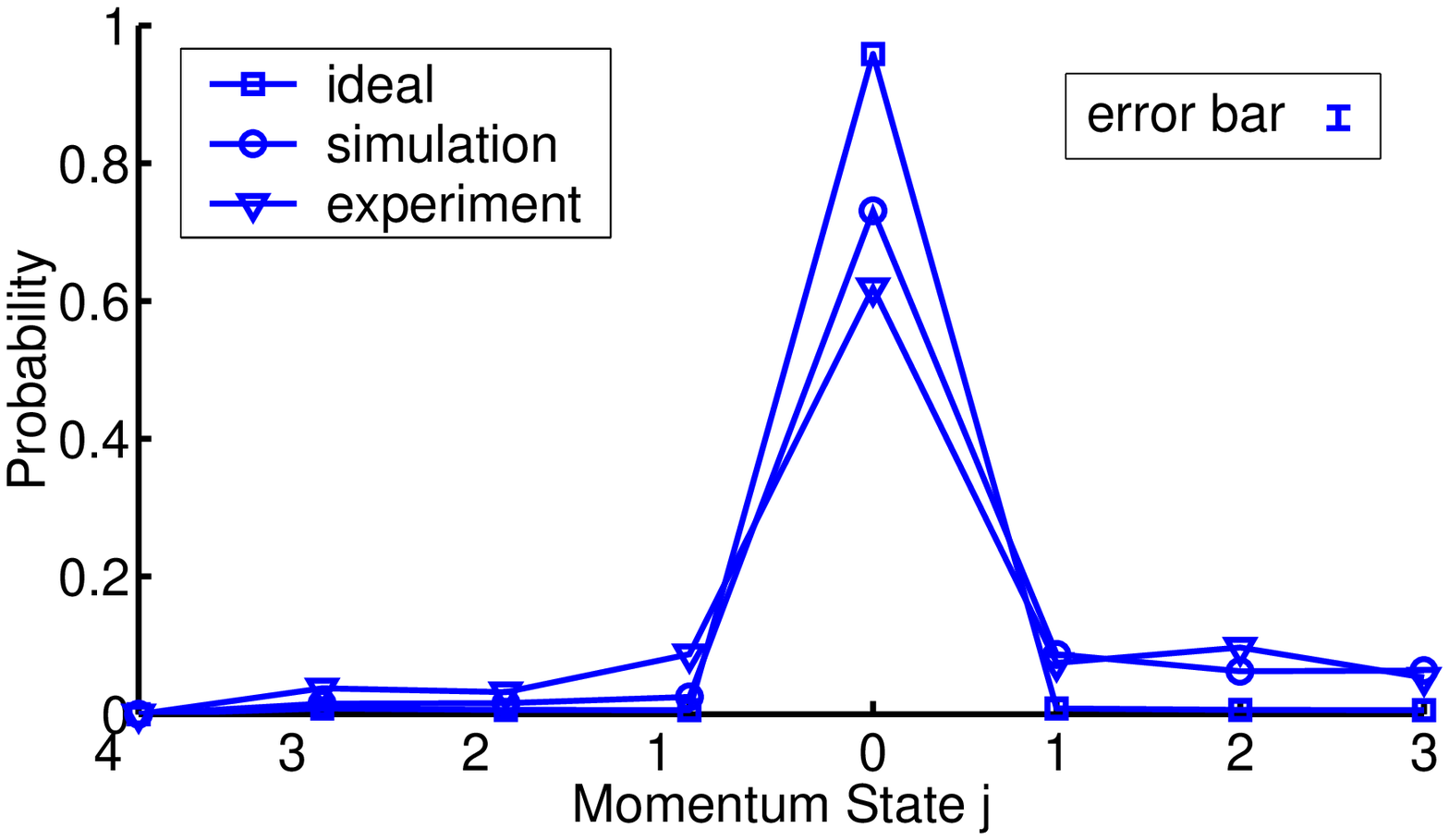}} & 
	\subfigure[1 iteration]{\includegraphics[width=3.0in]{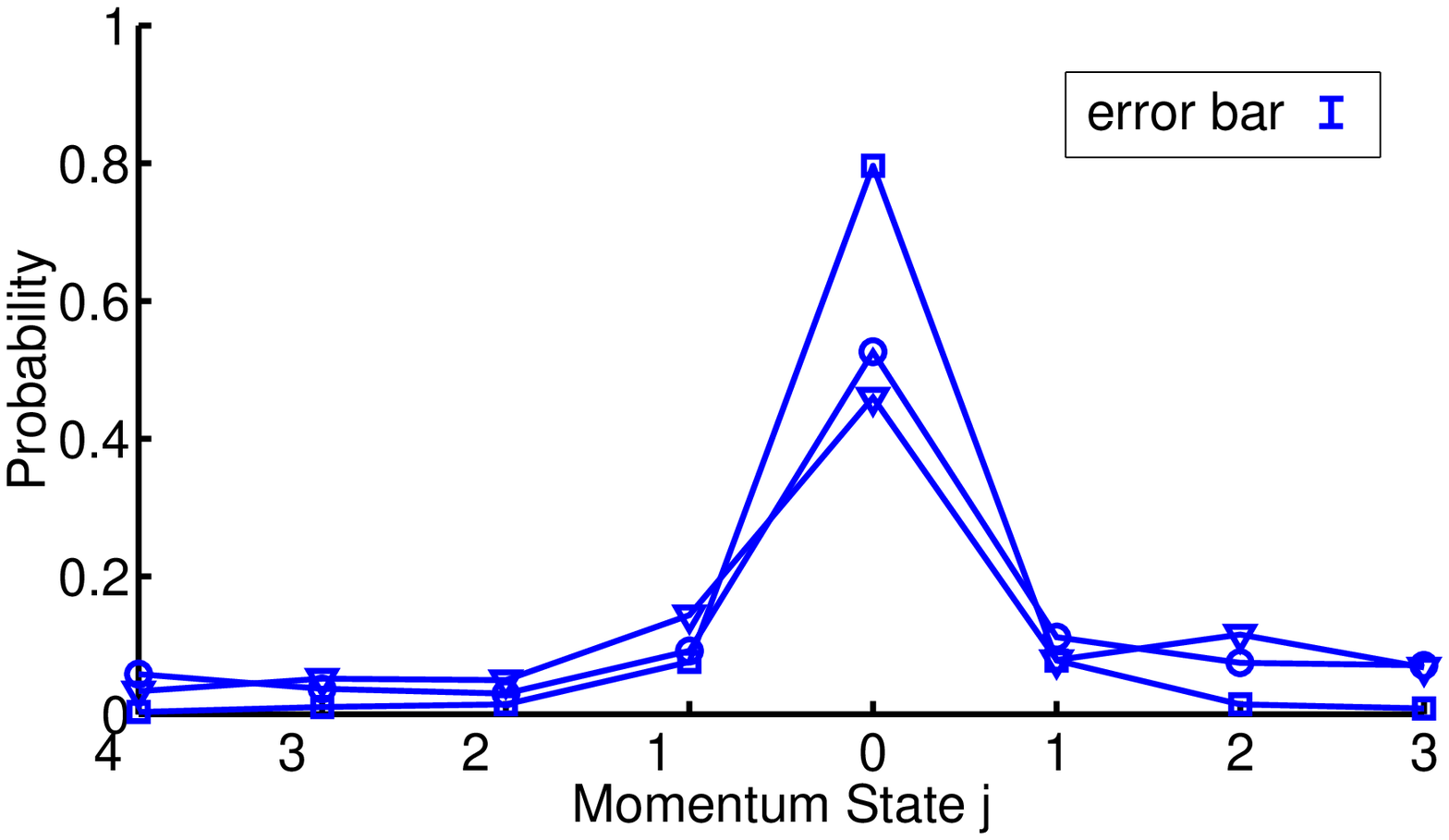}} \\
	\subfigure[2 iterations]{\includegraphics[width=3.0in]{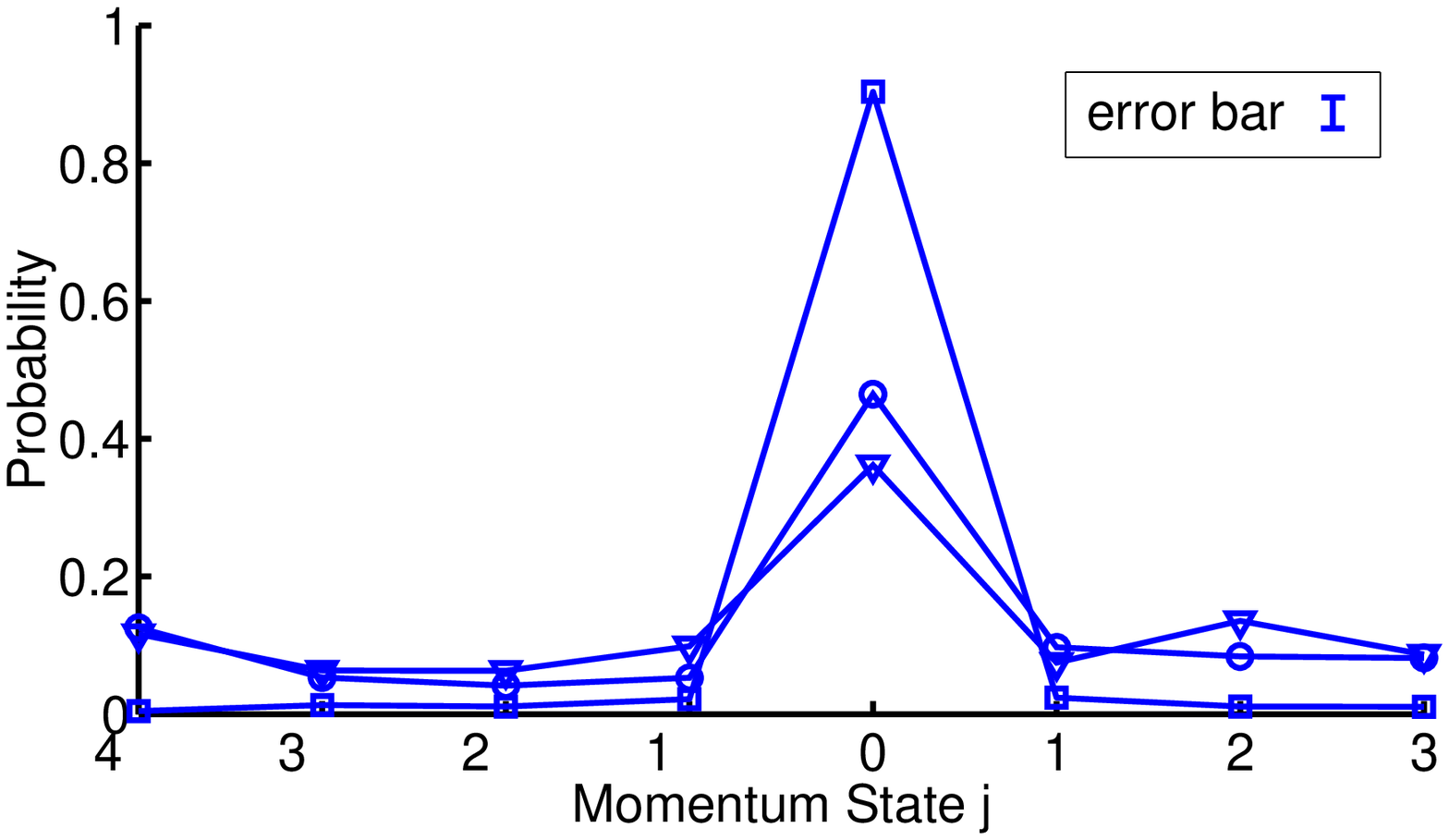}} &
	\subfigure[3 iterations]{\includegraphics[width=3.0in]{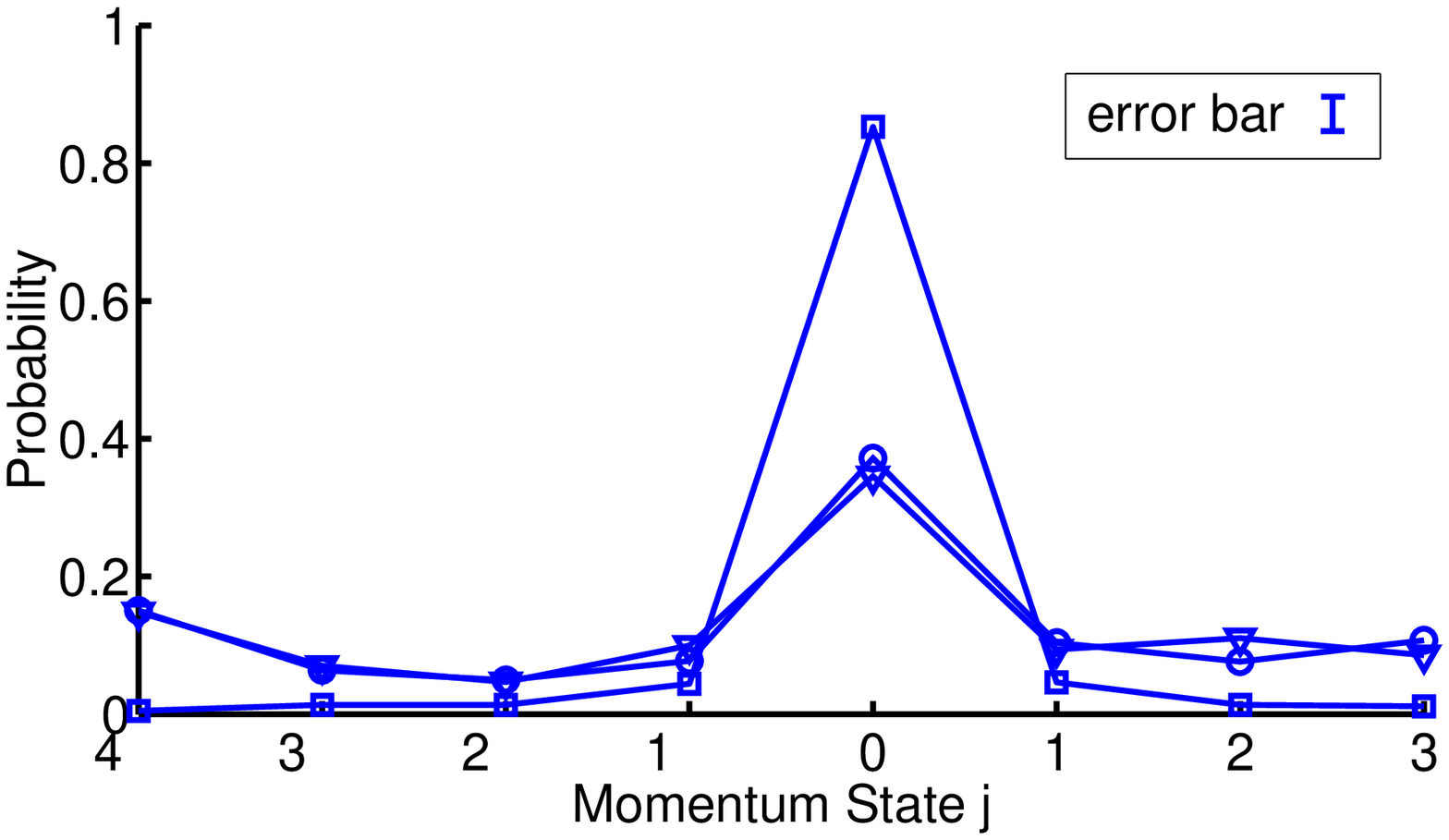}}
	\end{tabular}
	\end{center}
	\caption{The momentum distribution after 0 through 3 iterations of the quantum sawtooth map ($L=7$, $K=1.5$, $N=8$) for the experiment (triangles), a numerical simulation of the experiment which is not affected by noise (squares), and finally a numerical simulation of the experiment which includes coherent errors, decoherence effects, and incoherent errors due to rf inhomogeneity in both the carbon and hydrogen control fields (circles).  Error bars are drawn to scale for each iteration.  The population of the initial state ($\ms=0$) dominates the distribution through 3 iterations of the experiment, and the width of the central peak is essentially unchanged.  However, experimental noise clearly causes the ensemble-averaged state to delocalize through the appearance of a baseline offset over the momentum distribution.}
	\label{fig:bytype_int}
\end{figure}

In Fig. \ref{fig:bytype_int}, the experimentally measured probability distributions after zero through three iterations of the quantum sawtooth map are plotted along with the ideal distributions and the distributions obtained by numerical simulations of the experiment which account for decoherence and the full two-dimensional distribution of rf powers. 
The experimental data reveals that the interior region of the momentum distribution does not broaden, as in a diffusive regime, but rather, the peak maintains roughly the same breadth, as predicted by simulations.  Meanwhile, the increasing background probability offset reveals the presence of imperfections in the implemented map, representing those regions of the ensemble which are not localized due to incoherence.  These qualitative features of the experimental data are reflected in the quantitative measures of localization discussed later.     

Discrepancies between the ideal and experimentally observed behavior are caused by experimental noise and decoherence influencing the implementation of quantum sawtooth map, in addition to imperfections in the experimentally prepared input state and in the readout steps.  Figure \ref{fig:bytype_int} reveals, on a qualitative level, that these discrepancies are well accounted for by the noise model used in numerical simulations.  The relative contribution of the distinct noise mechanisms to the experimentally observed delocalization of the state is discussed further Sec. \ref{sec:discussion}.
  
\begin{figure}
	\begin{center}
  \includegraphics[width=3.2in]{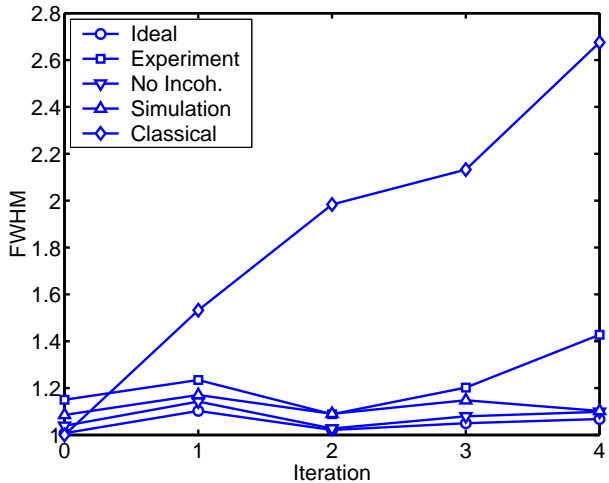}
	\end{center}
	\caption{The full width at half maximum of the momentum distribution after zero through four iterations of the sawtooth map in various implementations:  a numerical simulation of the exact classical map (diamonds), a numerical simulation of the exact quantum map (circles), the experimentally implemented map (squares), a numerical simulation of the experimentally implemented map which accounts for decoherence without incoherence (triangles down), a numerical simulation of the experimentally implemented map which accounts for decoherence with incoherence (triangles up).  The FWHM reveals that despite the noise affecting the QIP, the distribution mimics the ideal quantum behavior, and does not broaden in a diffusive manner as in the classical case.}
	\label{fig:fwhm}
\end{figure}

In light of the incoherent variations of localization properties of the map over the ensemble, we wish to select a measure that can be interpreted as the extent to which some portion of the ensemble demonstrates quantum localization.  By measuring the full width at half maximum (FWHM) of the probability distribution in successive iterations, we can observe the presence of any dynamical broadening of the distribution, without regard to the background probability offset caused by incoherence, as discussed in Sec. \ref{sec:num_sim}.   Fig. \ref{fig:fwhm} shows a plot of the FWHM for each probability distribution plotted in Fig. \ref{fig:bytype_int}.  
The FWHM data reveals the dynamical properties of the experimentally measured distribution as distinct from the classical behavior.  The relative flatness of the experimentally measured FWHM curve through three iterations of the map is consistent with quantum localization in an incoherent ensemble.  Numerical simulations show the progression in peak width from the ideal simulation (most narrow) to simulations where decoherence and incoherence are included in the simulation.  The numerical simulation which accounts for decoherence but not incoherence corresponds to the momentum distributions plotted in Bin 6 of Fig. \ref{fig:bins}.  

\section{Discussion}  
\label{sec:discussion}

\begin{figure}
	\begin{center}
	\includegraphics[width=3.2in]{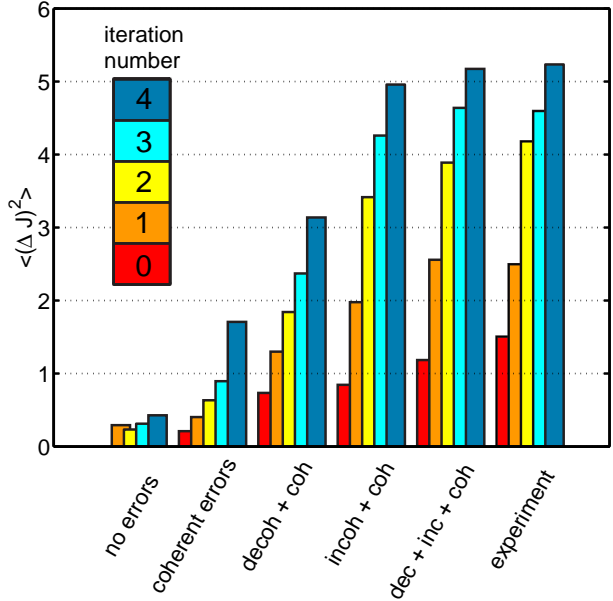}
	\end{center}
	\caption{The second moment of the probability distribution determined from numerical simulations of the experiment including the error models discussed in the text, compared to the ideal data and the experimental data.  This plot demonstrates the relative importance of the individual noise mechanisms as they contribute to the experimentally observed delocalization process.  As more errors are included in numerical simulations, the system shows stronger delocalization and more closely emulates the experimental data.}
	\label{fig:sm_full}
\end{figure}

By using numerical simulations to isolate the various types of errors known to influence the experiment, it is possible to measure the relative significance of each type of error by examining the degree to which it leads to delocalization in the system.  Figure \ref{fig:sm_full} shows the degree to which each type of error causes delocalization in the resulting state, as measured by the second moment of the corresponding probability distribution, thus distinguishing the relative importance of the distinct noise mechanisms in the experiment.
The data also reveals the extent to which the breadth of the distribution is affected by errors in the initial state preparation.  Evidently, the coherent errors are essentially inconsequential over at least three iterations of the map.  The slope of second moment versus time plot is most strongly affected by incoherent errors, and thus incoherence is determined to be the dominant noise mechanism limiting the degree to which localization is acheived by experimental control, which is consistent with the observations of Sec. \ref{sec:num_sim}.  

\begin{figure}
	\begin{tabular}{ccccc}
	\multicolumn{2}{c}{1st Krauss Op.} &
	        &
	\multicolumn{2}{c}{Superoperator} \\
	Iter. 1 &
	Iter. 2 &
	        &
	Iter. 1 &
	Iter. 2 \\
	\includegraphics[width=0.7in]{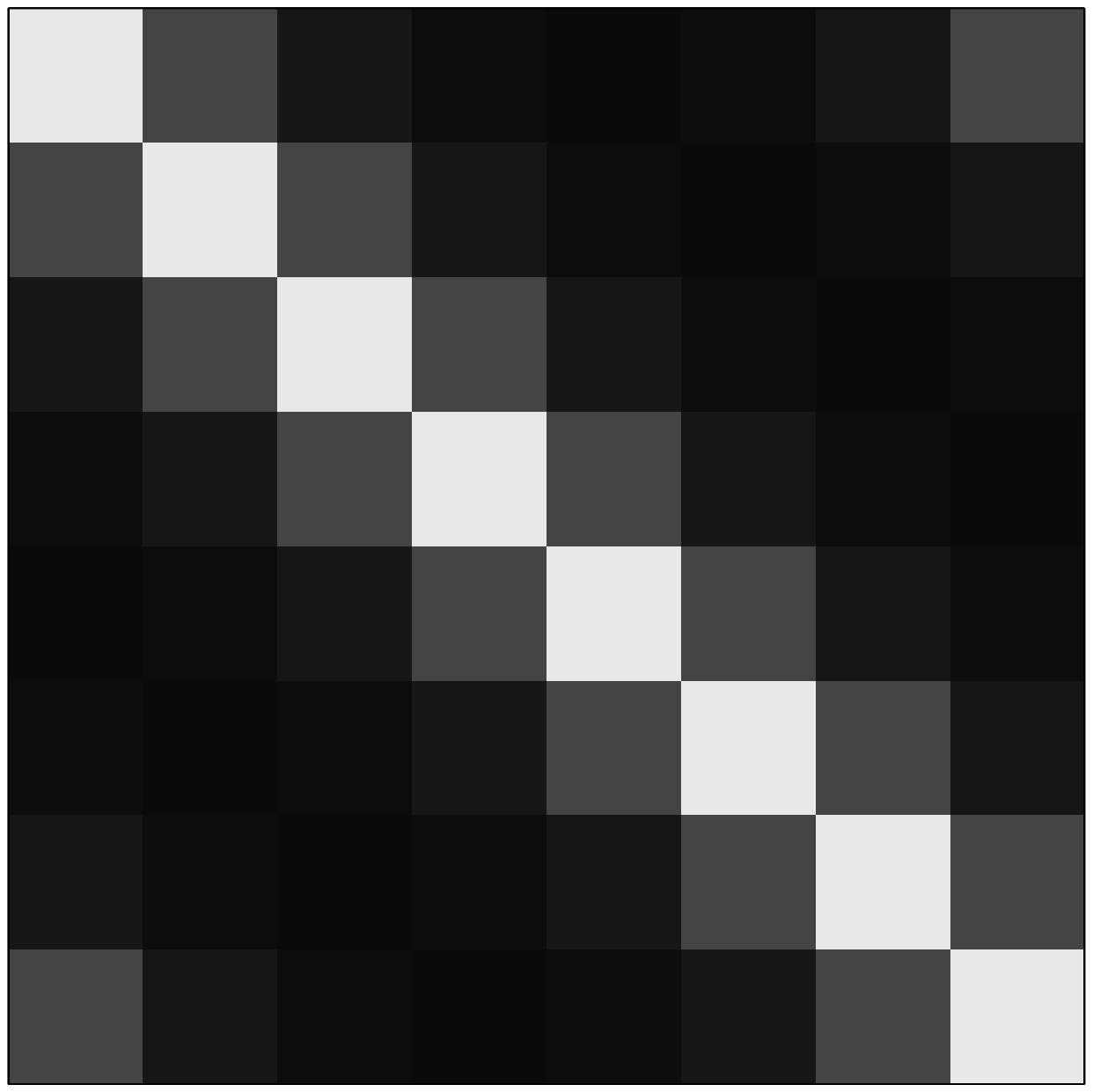} &
	\includegraphics[width=0.7in]{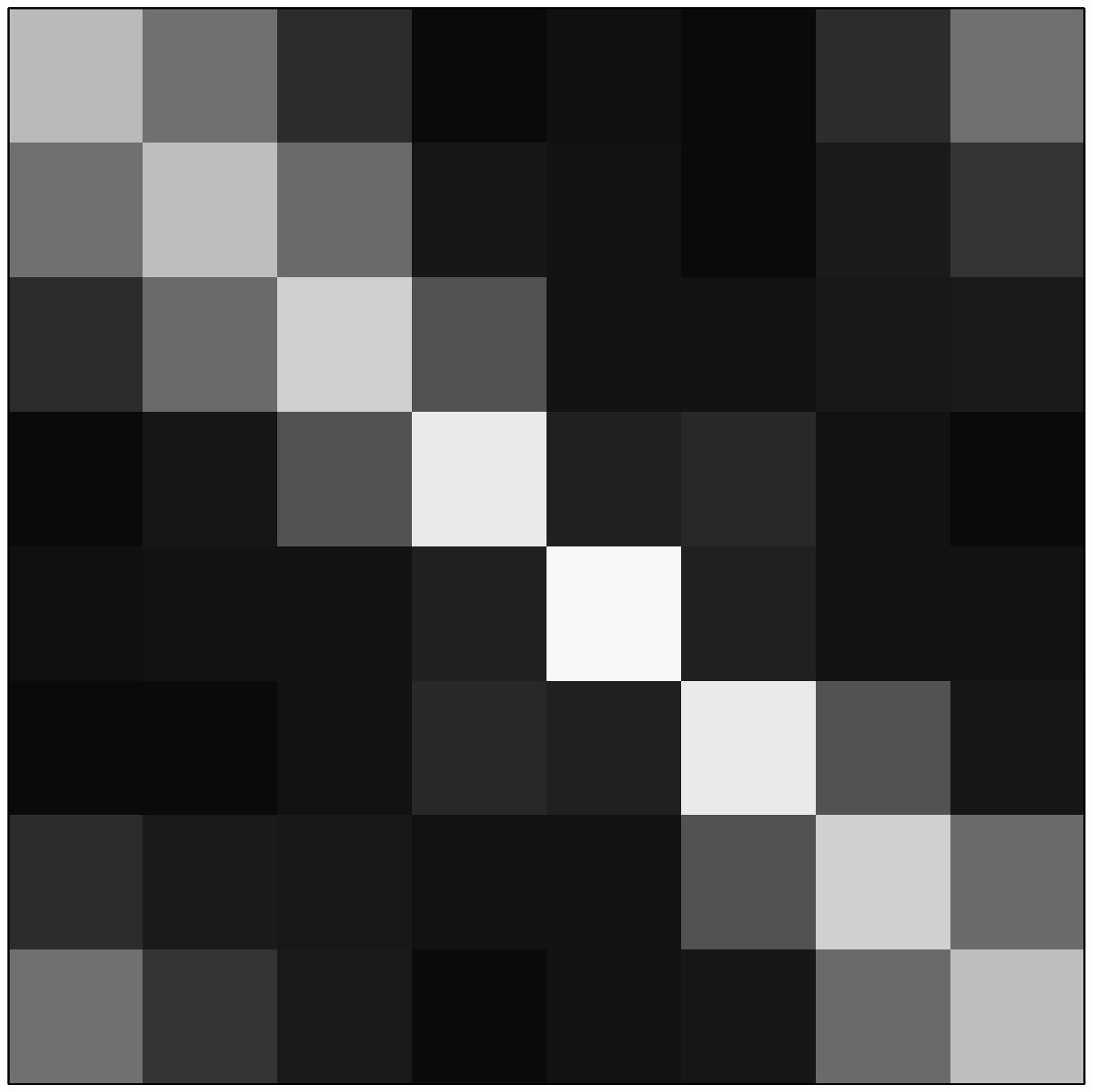} &
	        &
	\includegraphics[width=0.7in]{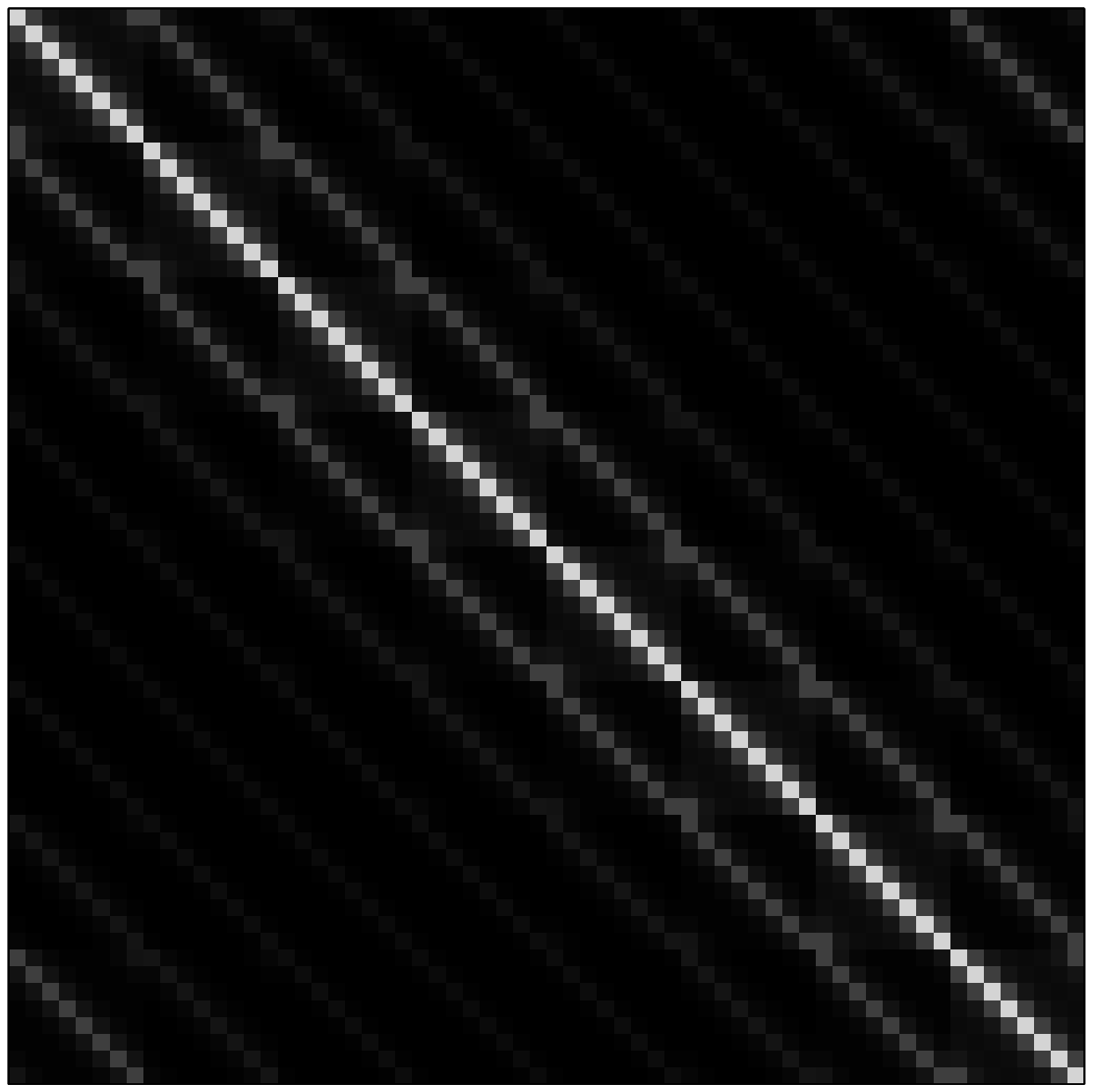} &
	\includegraphics[width=0.7in]{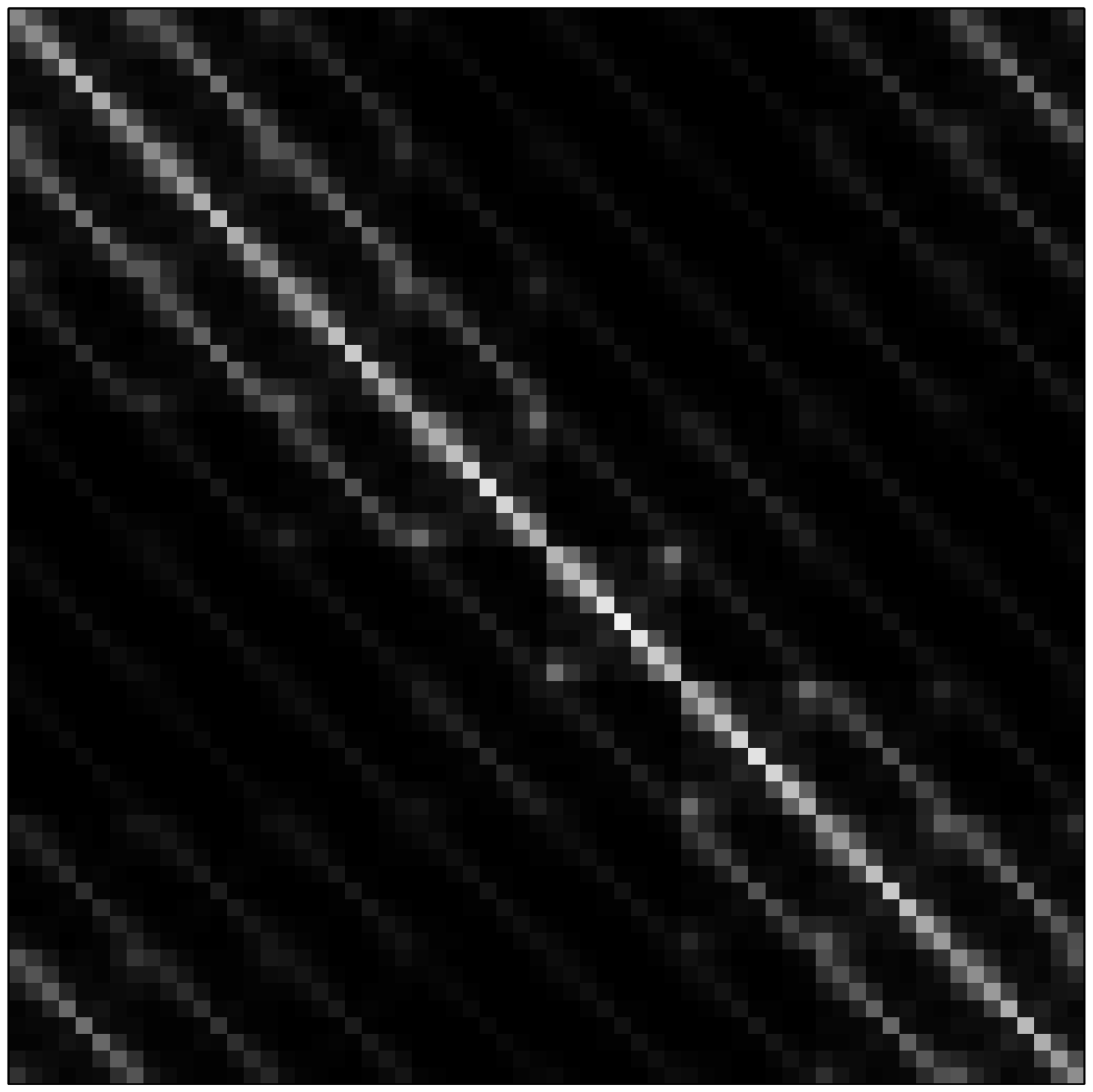} \\
	\includegraphics[width=0.7in]{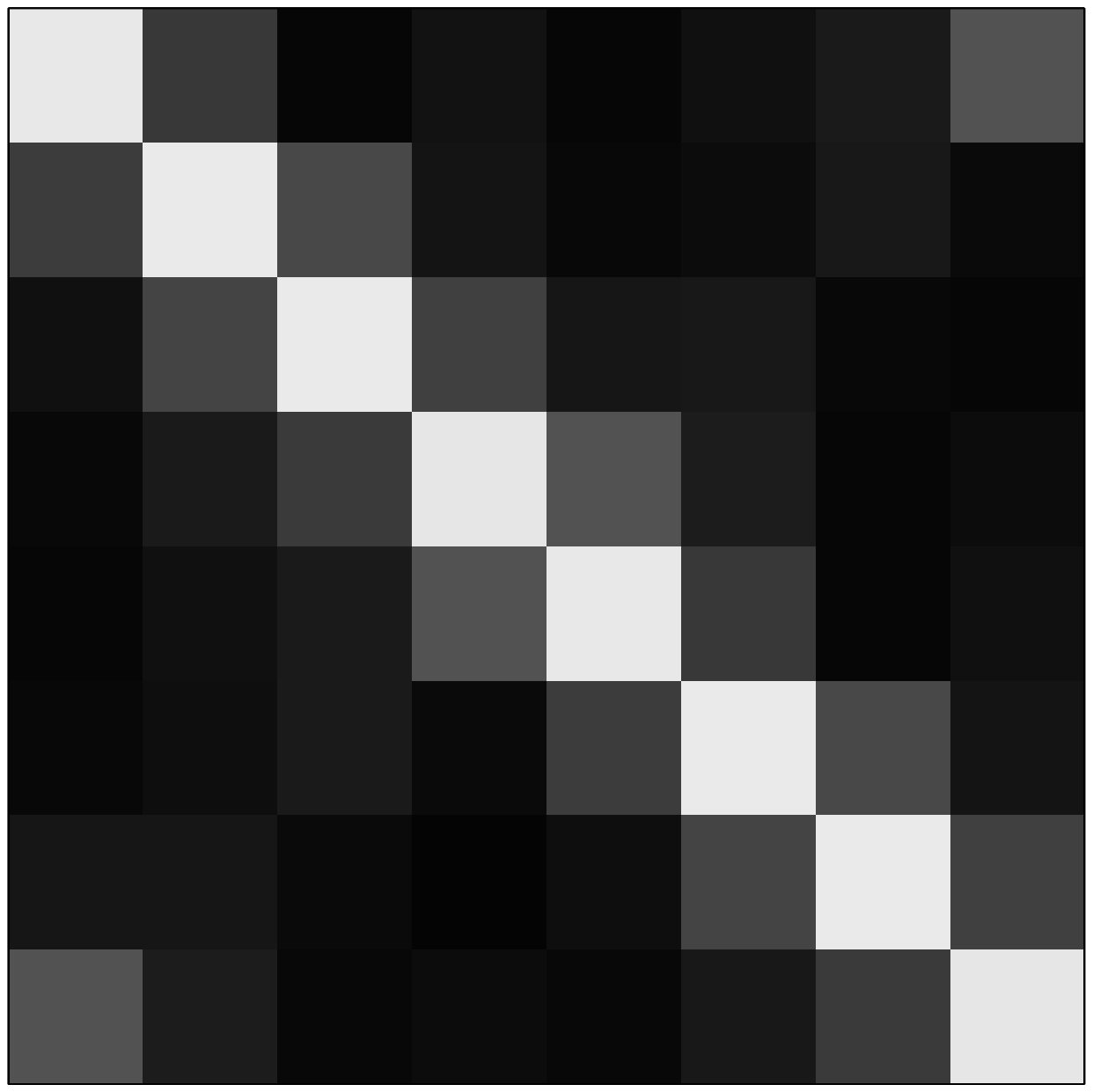} &
	\includegraphics[width=0.7in]{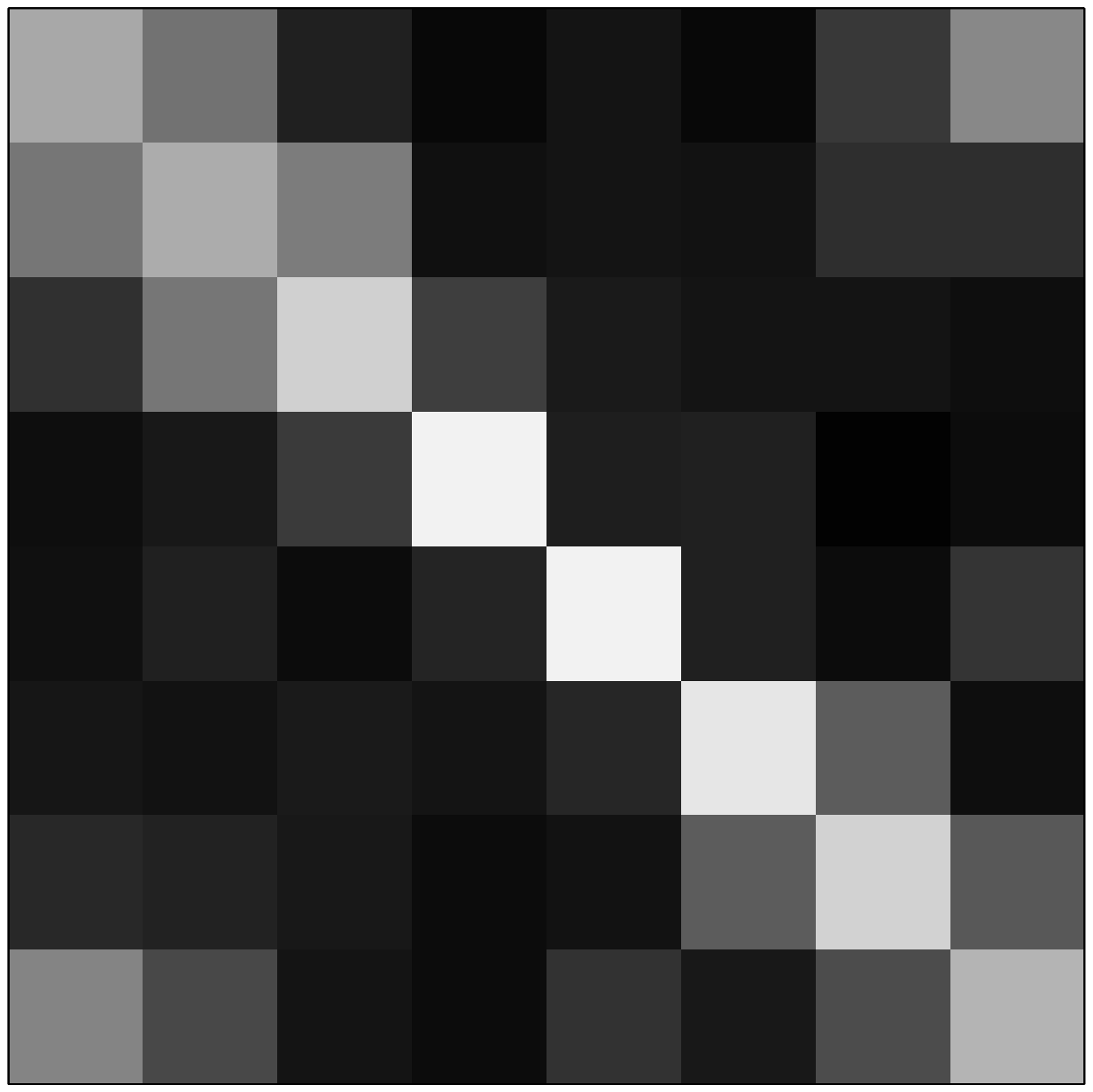} &
	        &
	\includegraphics[width=0.7in]{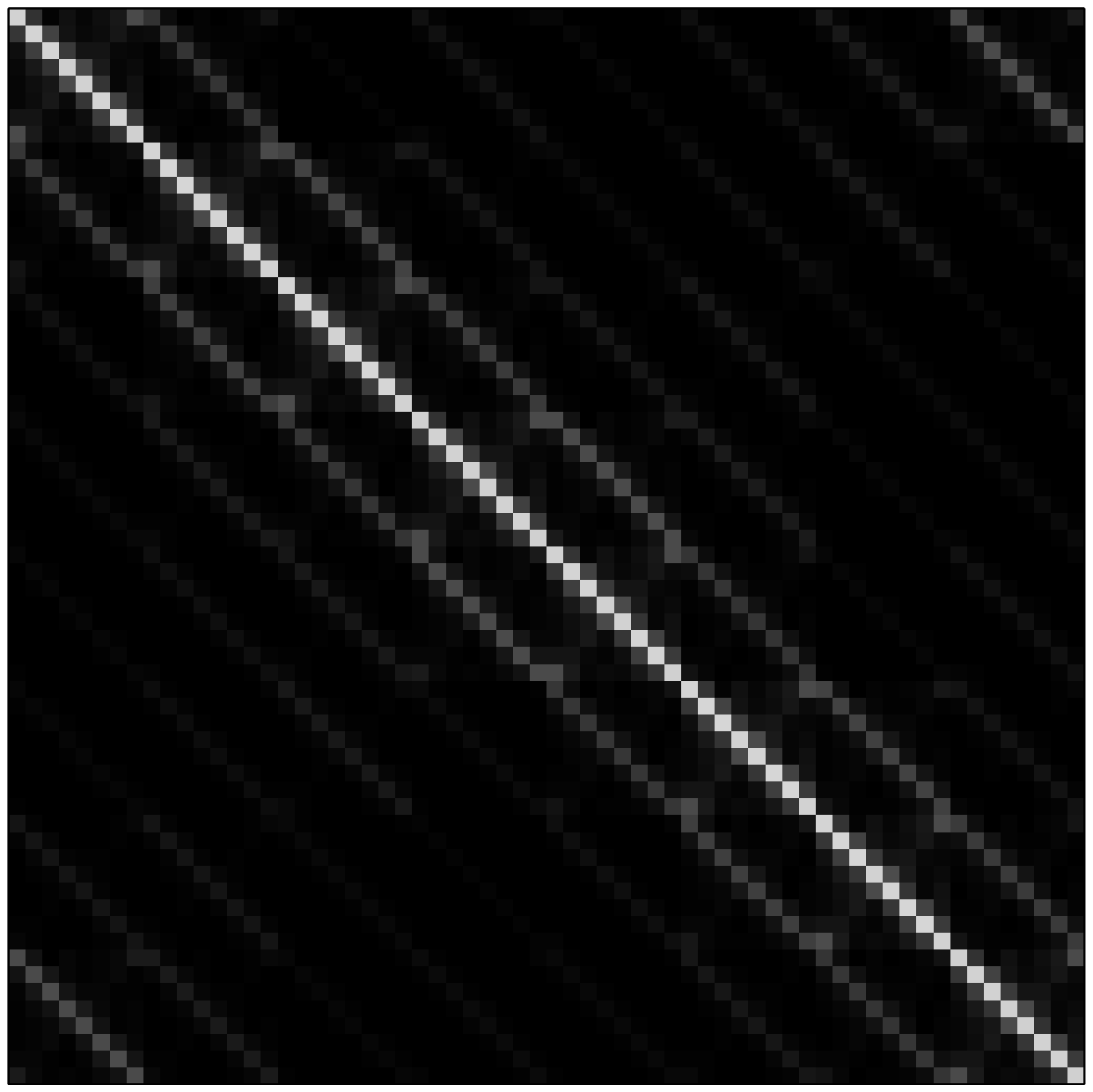} &
	\includegraphics[width=0.7in]{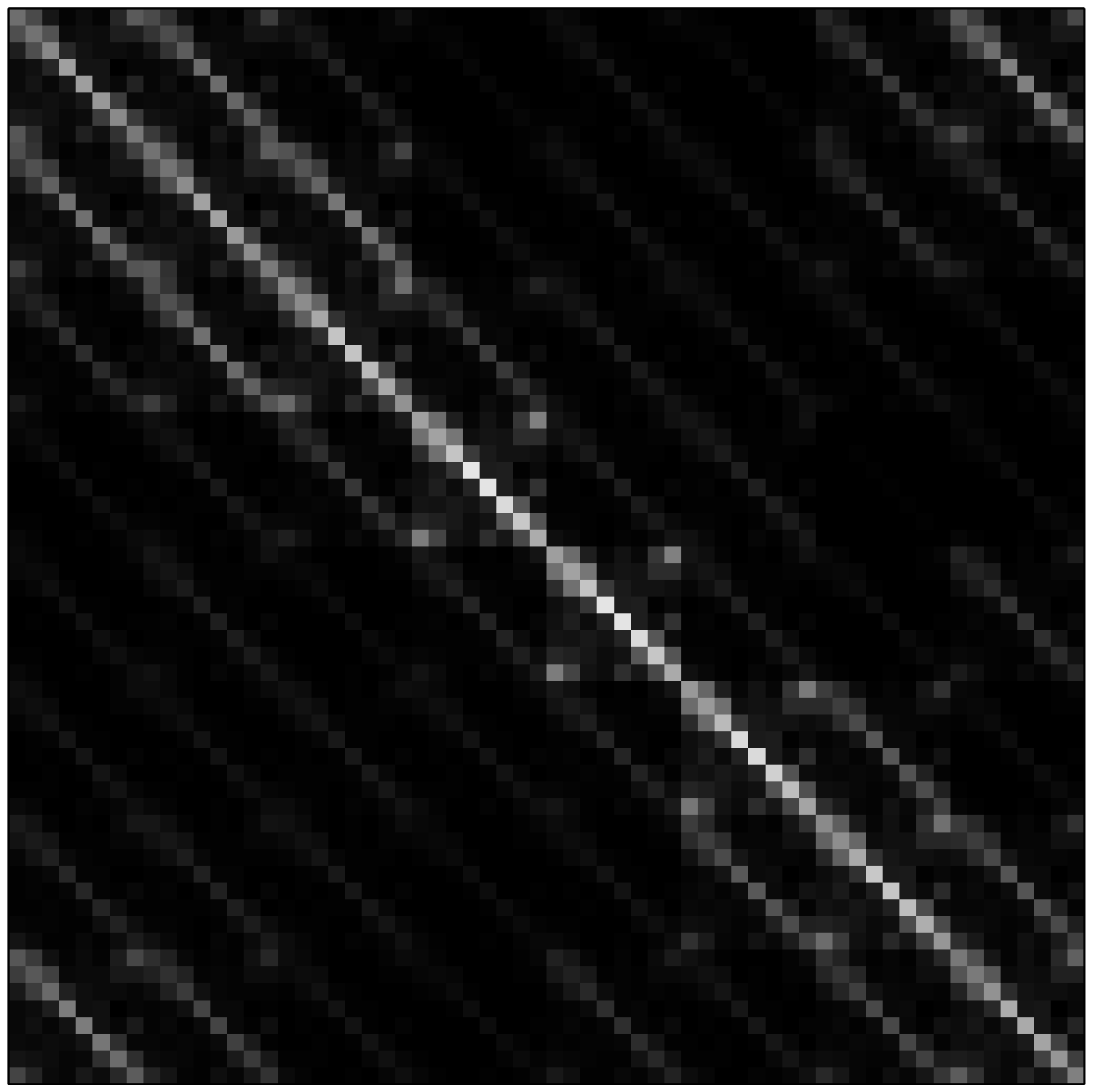} \\
	\includegraphics[width=0.7in]{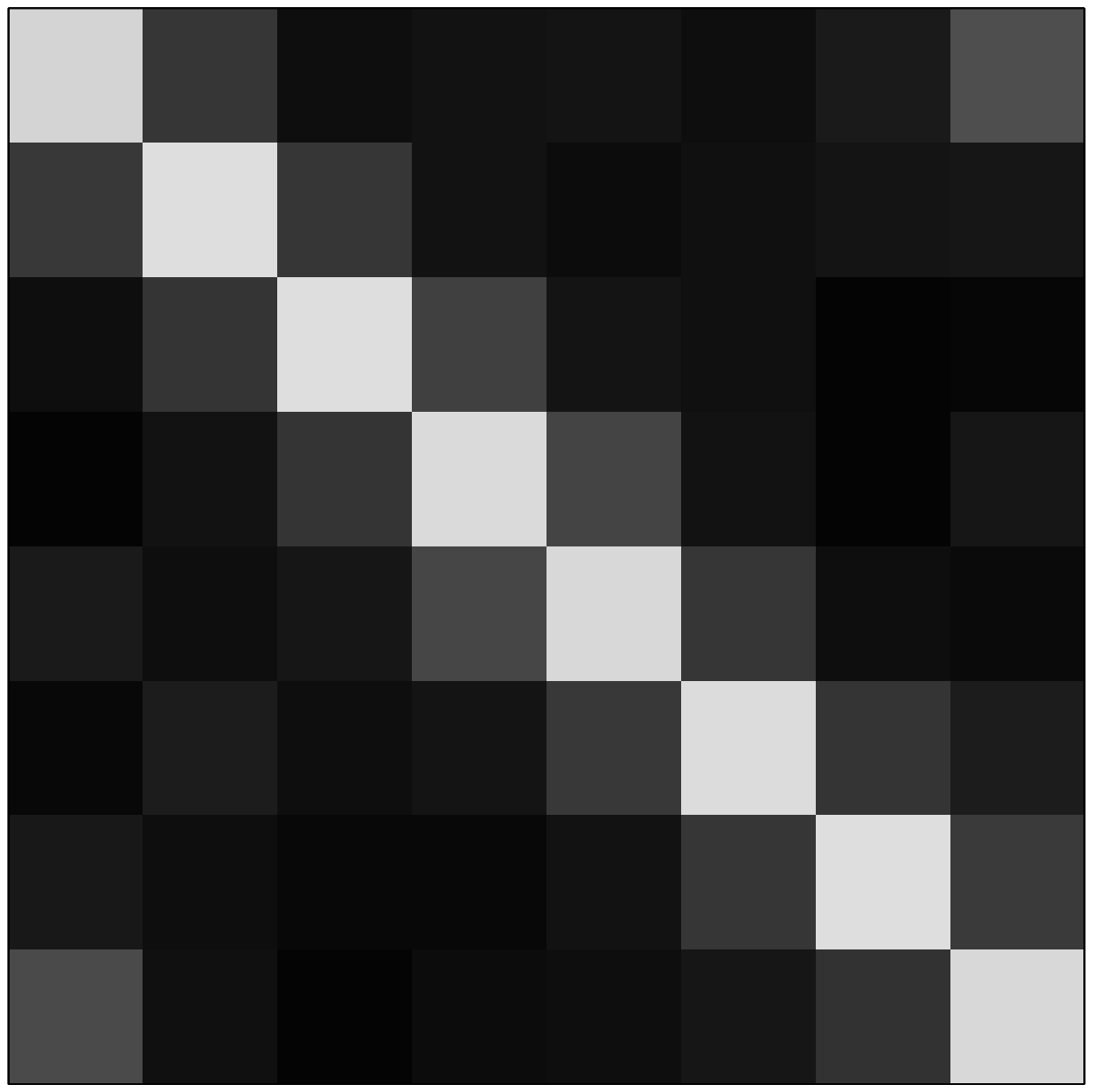} &
	\includegraphics[width=0.7in]{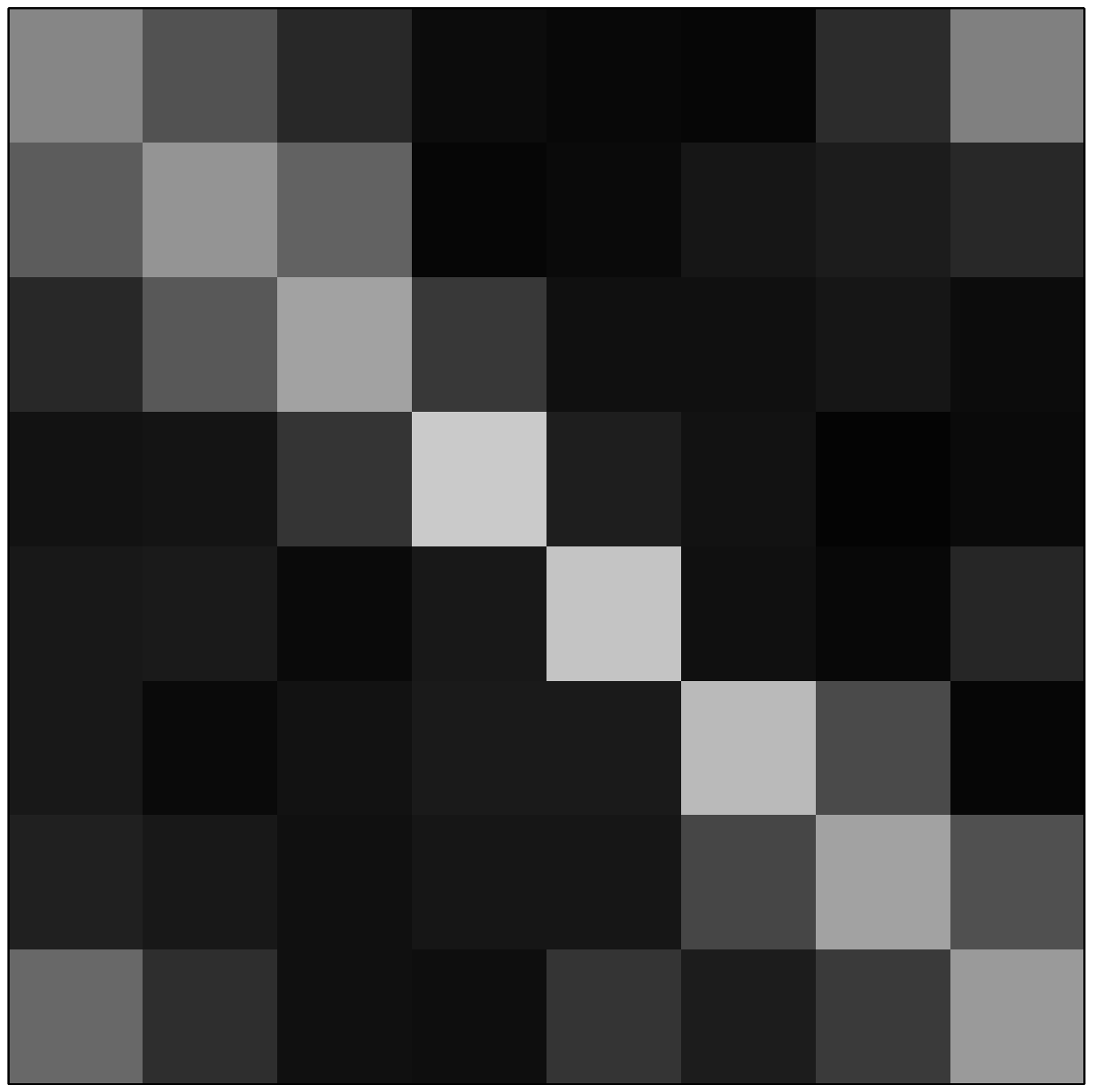} &
	        &
	\includegraphics[width=0.7in]{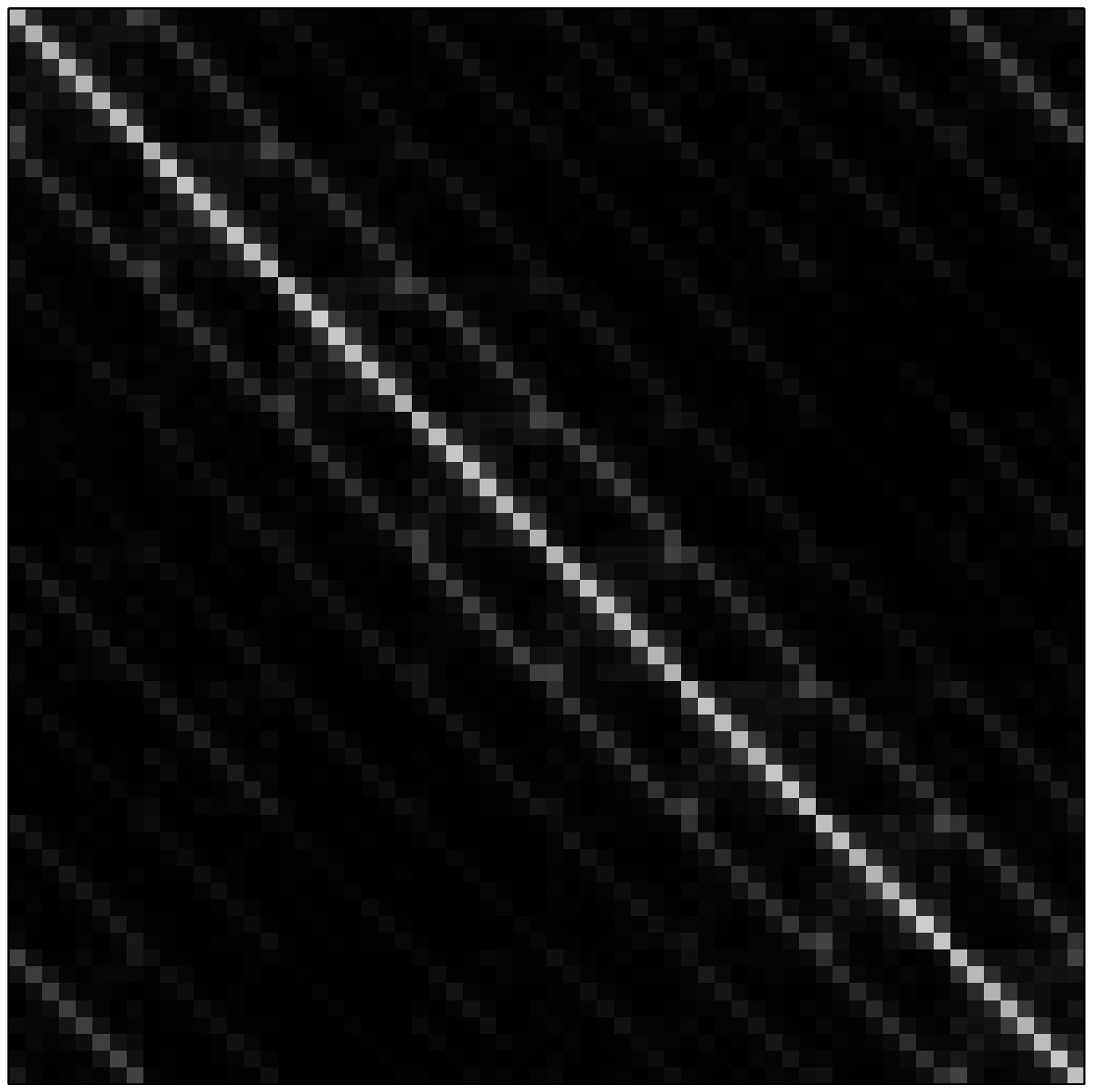} &
	\includegraphics[width=0.7in]{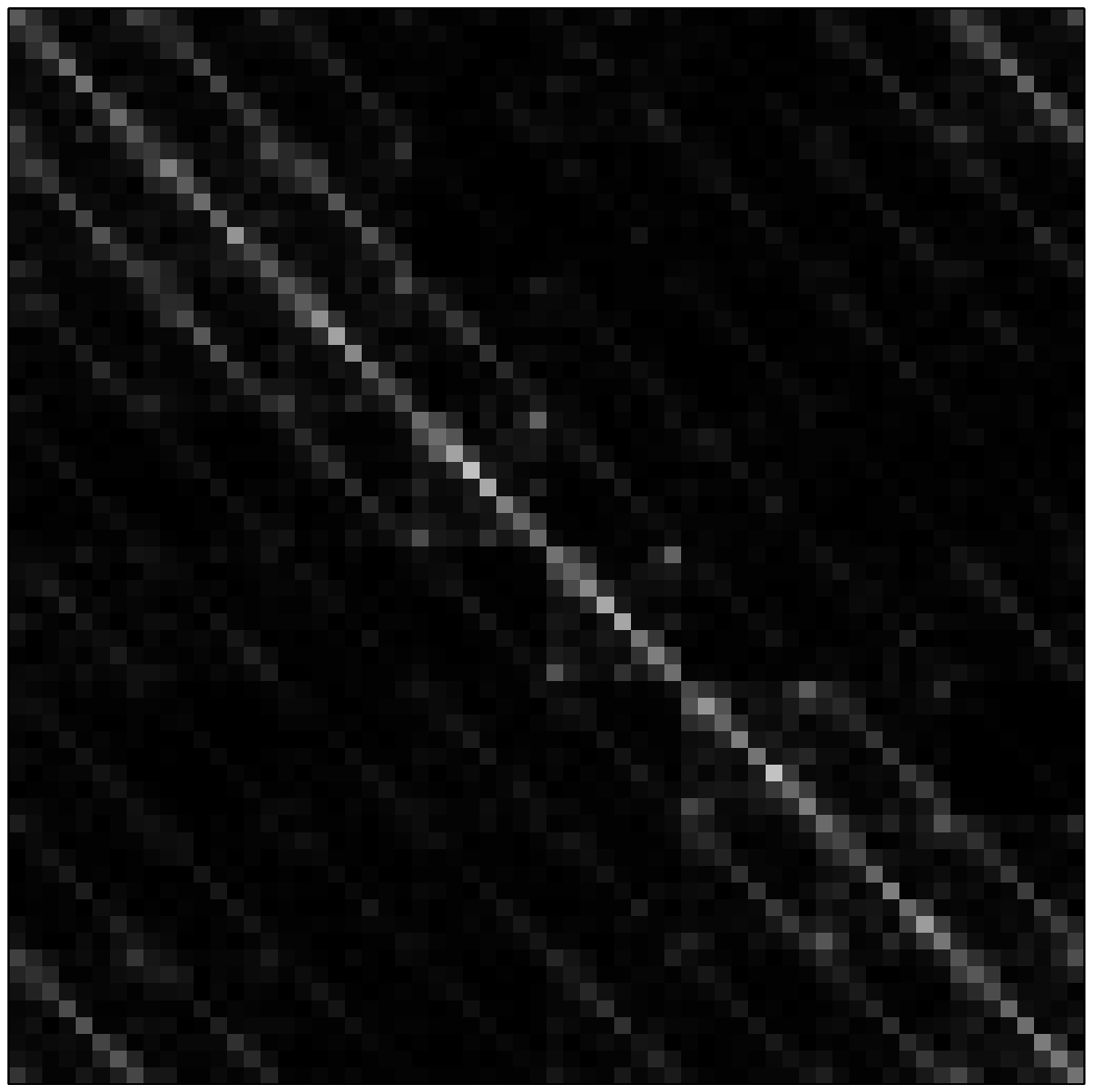} \\
	\includegraphics[width=0.7in]{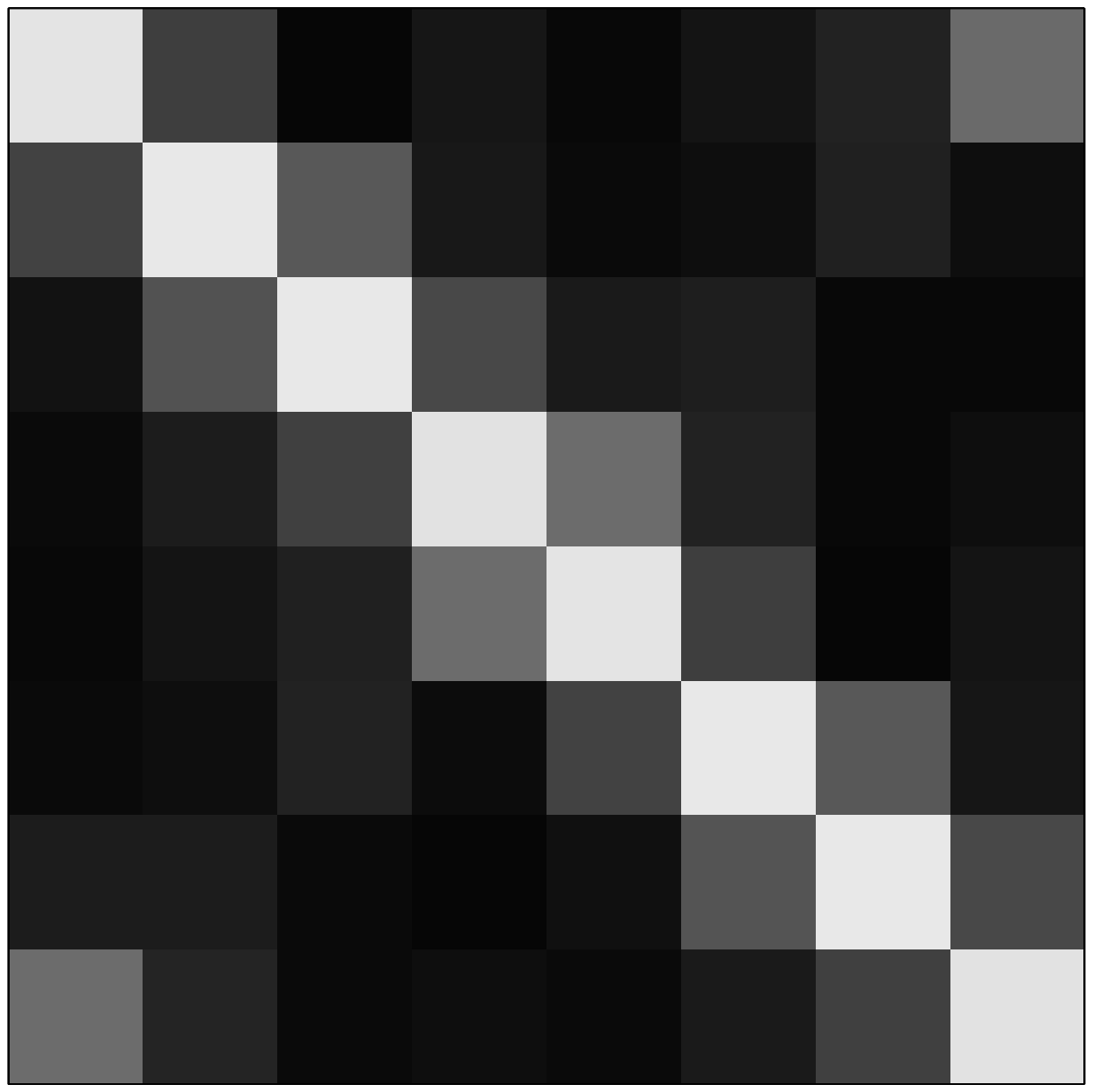} &
	\includegraphics[width=0.7in]{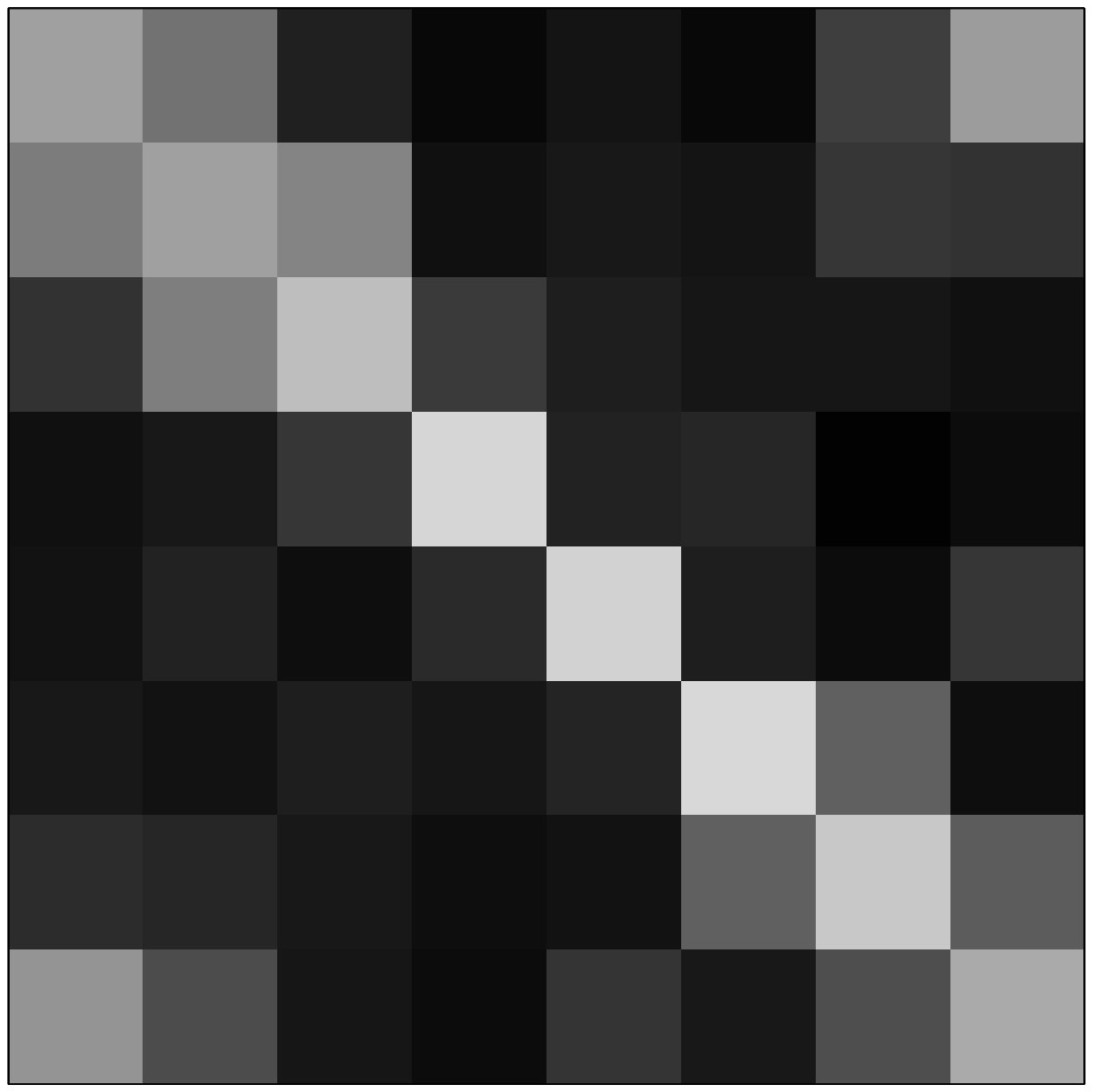} &
	        &
	\includegraphics[width=0.7in]{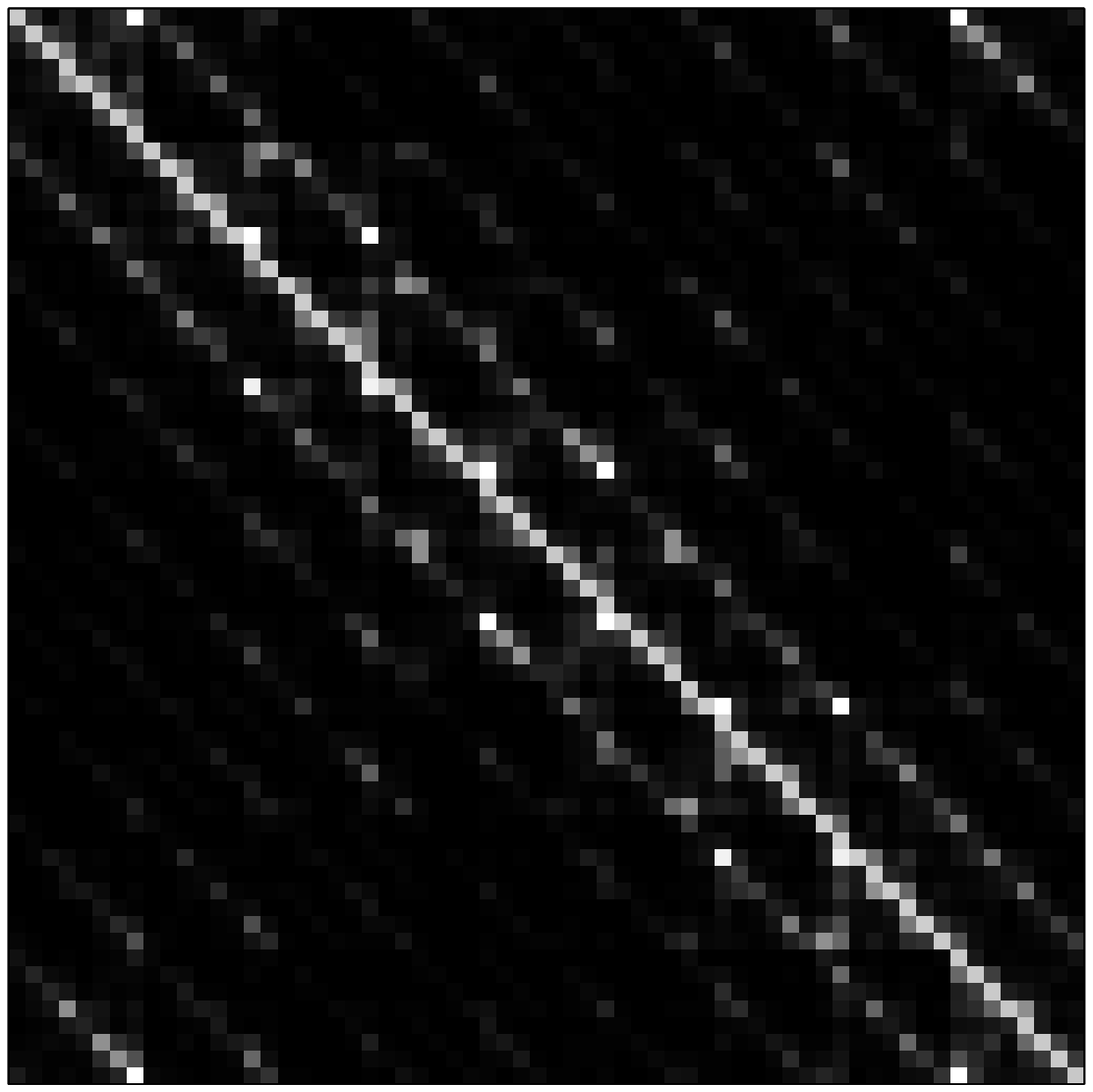} &
	\includegraphics[width=0.7in]{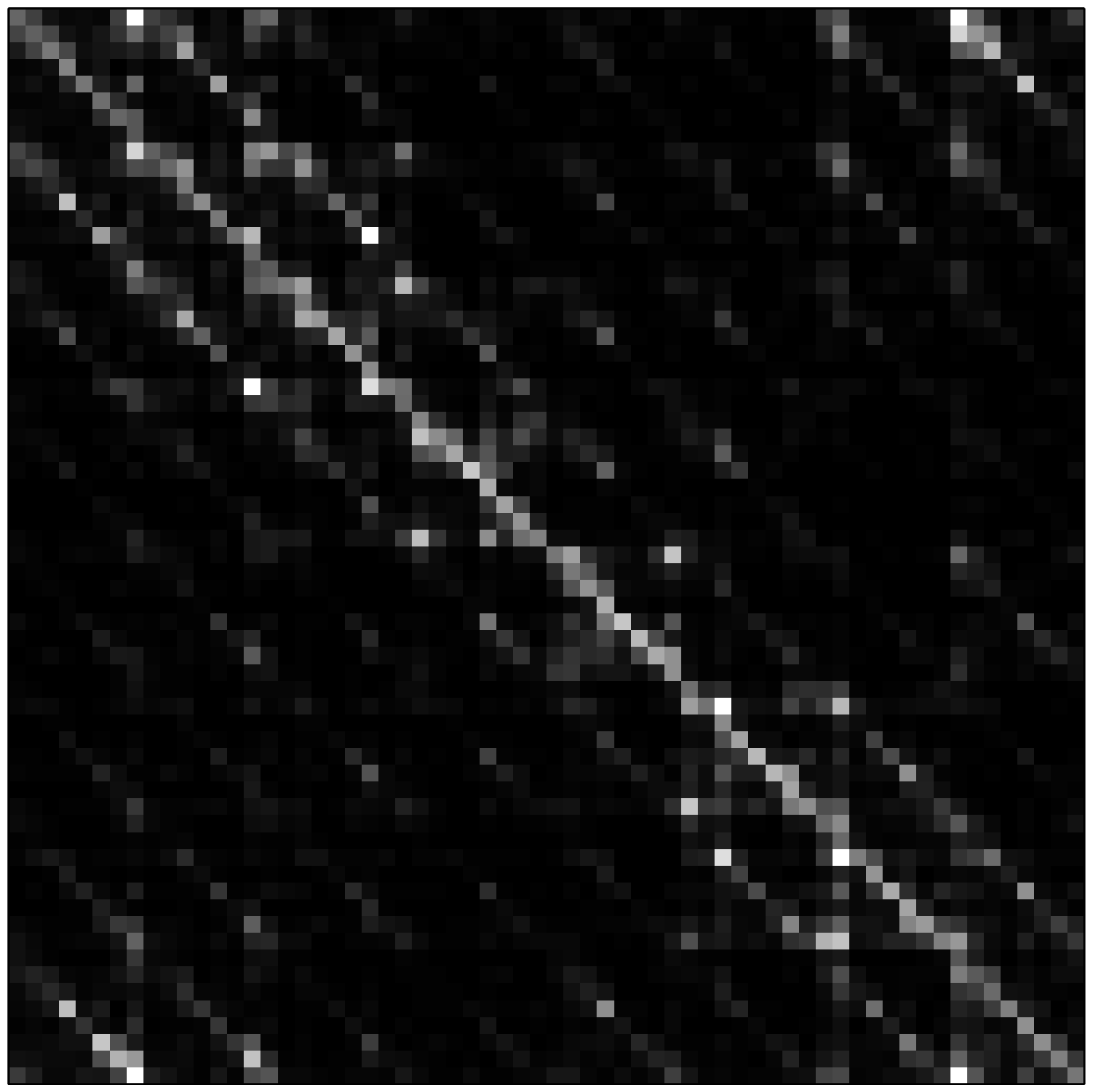} \\
	\includegraphics[width=0.7in]{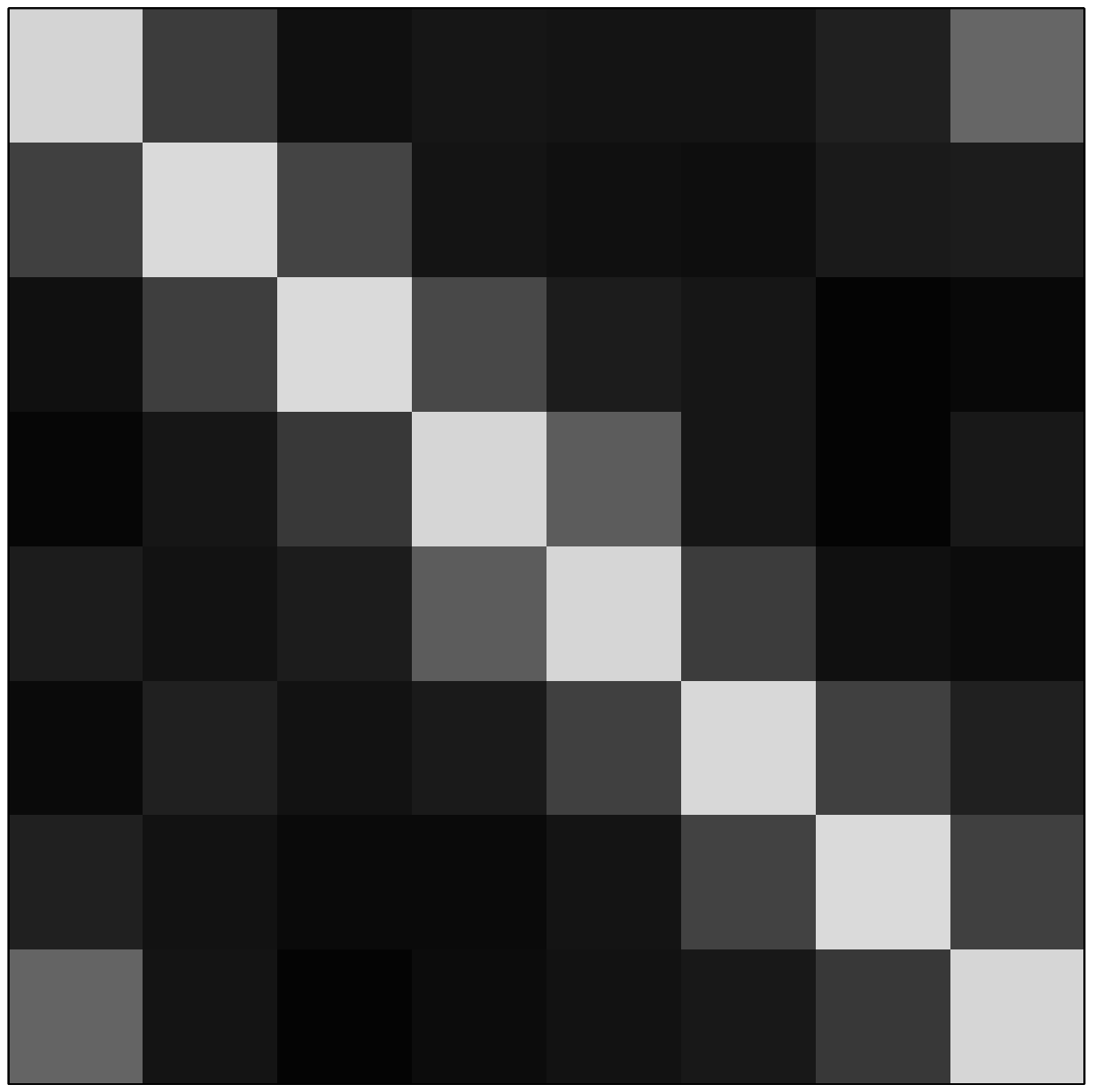} &
	\includegraphics[width=0.7in]{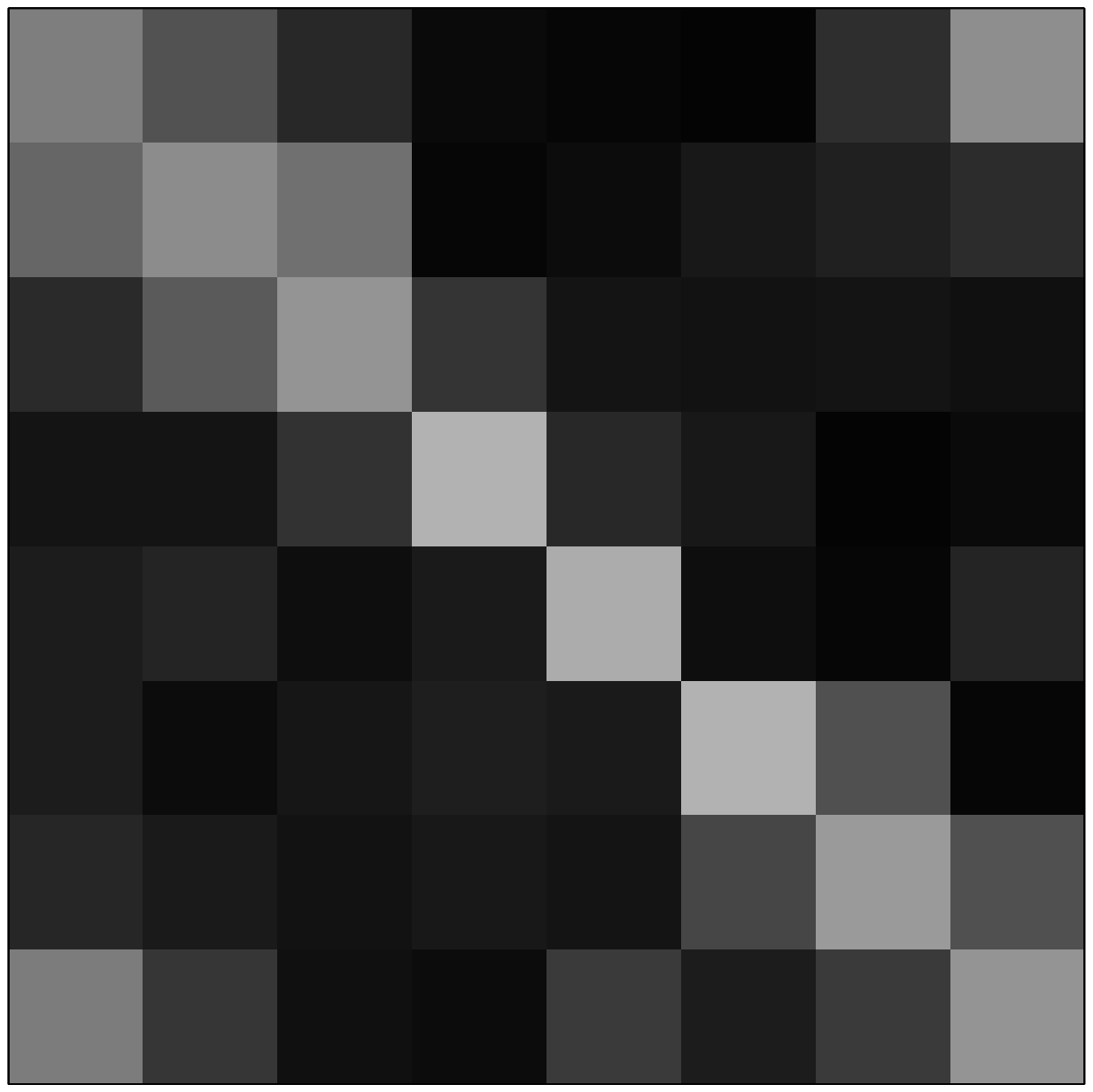} &
	        &
	\includegraphics[width=0.7in]{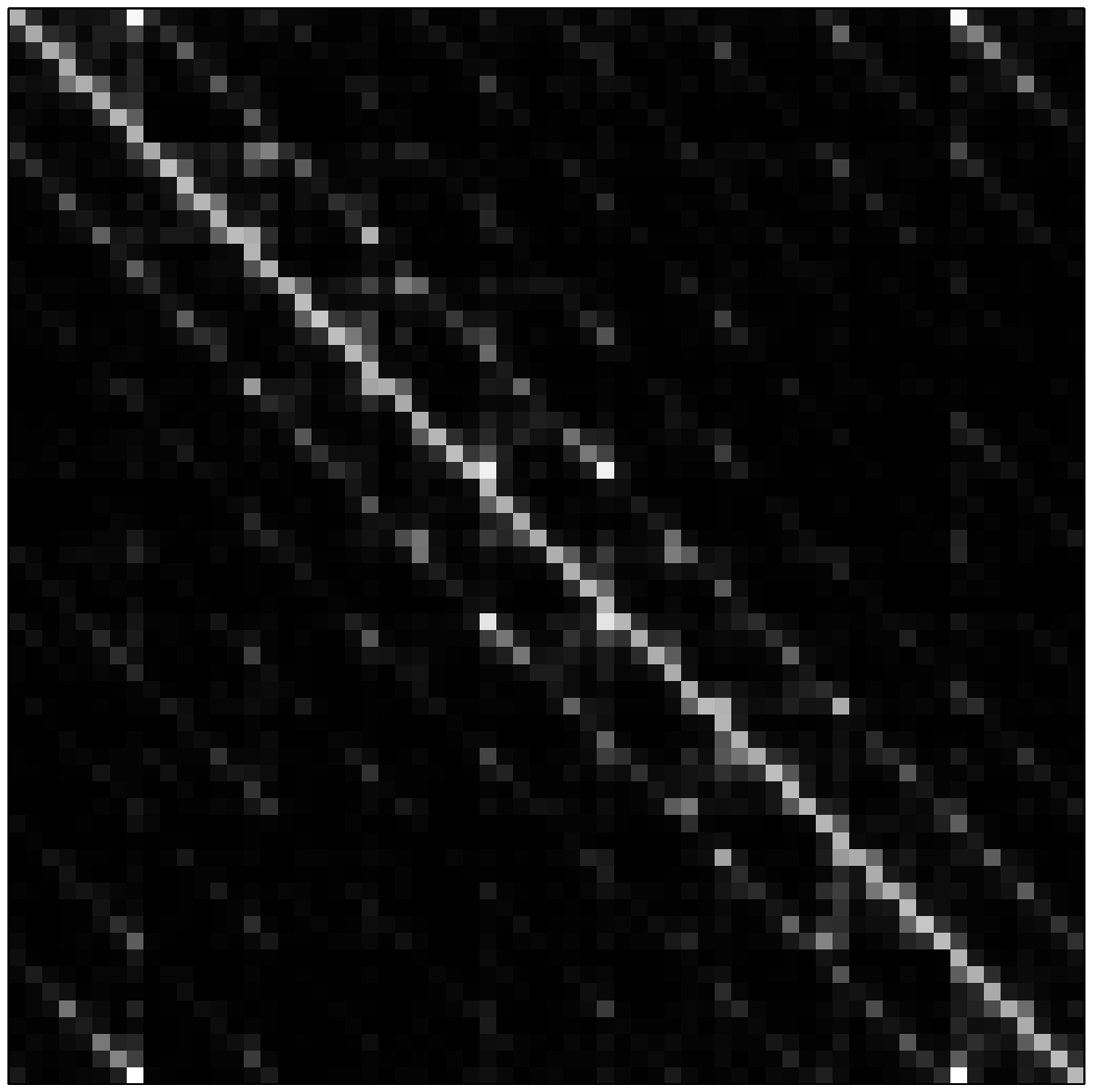} &
	\includegraphics[width=0.7in]{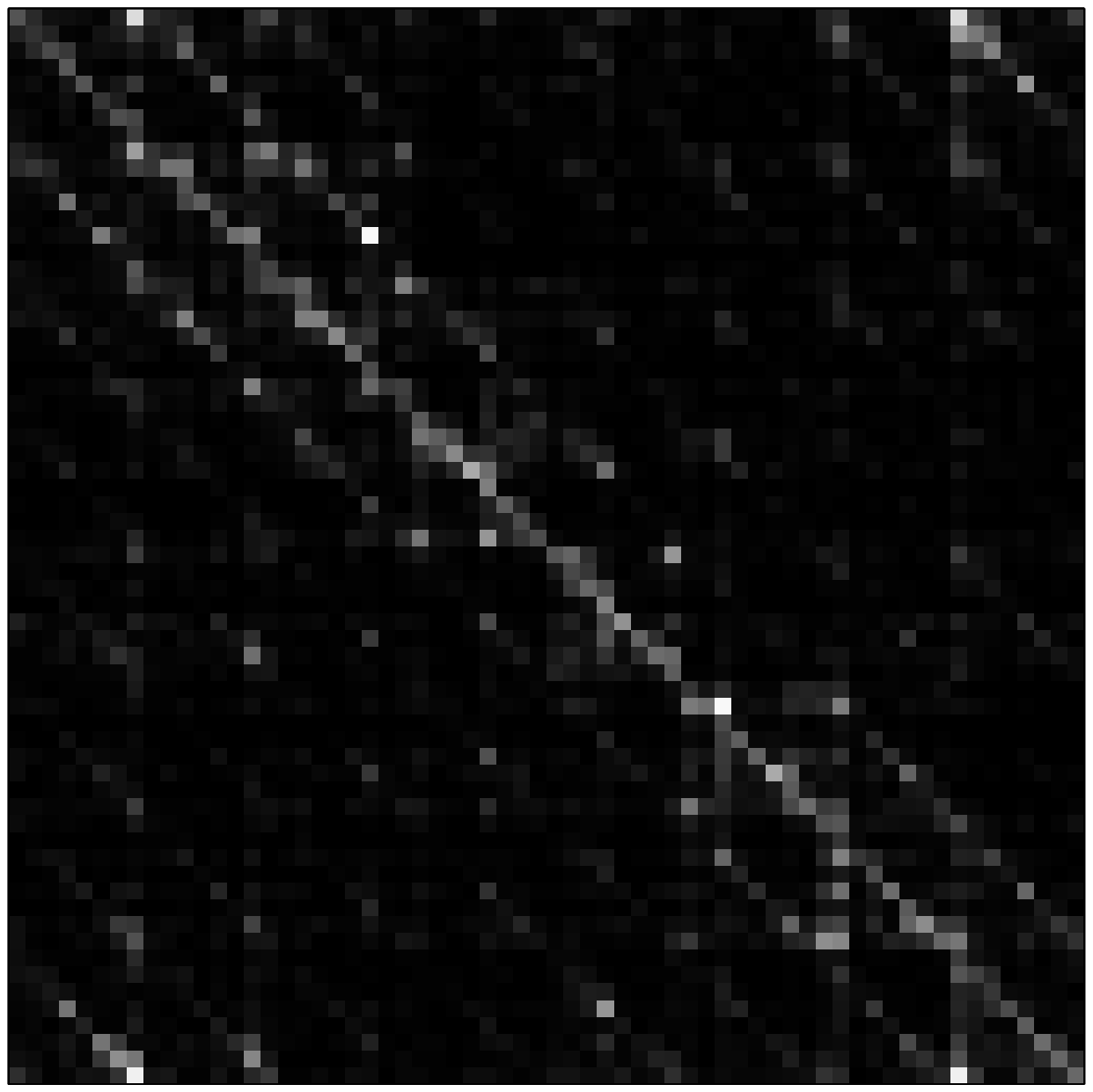} \\
	\end{tabular}
	\caption{The magnitude of each element of the superoperators and most significant Kraus operators for numerical simulations of the quantum sawtooth map experiment which include different types of errors (top to bottom): (1) no errors, (2) coherent errors, (3) coherent errors and decoherent errors due to relaxation, (4) coherent errors and incoherent errors due to rf inhomogeneity, (5) coherent, incoherent, and decoherent errors.  The elements are scaled from zero (black) to one (white).  The errors in the implementation of the sawtooth map affect the bandedness of each Krauss operator (and similarly of each superoperator).}
	\label{fig:sup_ops}
\end{figure}

Additional insight on the delocalizing effects of experimental noise and decoherence can be gained by examining the superoperators and the corresponding Kraus operators for each type of numerical simulation.  A superoperator of dimension $N^2 \times N^2$ which describes a completely positive quantum process can be equivalently expressed as an operator sum, which involves at most $N$ Kraus operators of dimension $N \times N$ \cite{Kraus1983}.  That is to say that for a general quantum process as in Eq. \ref{eq:supopevol},
\begin{equation}
\DvR{\rho_{out}}=S\DvR{\rho_{in}}
\end{equation}
there is an equivalent representation of the form
\begin{equation}
\rho_{out}=\sum_kA_k\rho_{in}A_k^{\dagger}
\end{equation}
where $A_k$ is the $k^{th}$ Kraus operator, which has a magnitude of $\left\|A_k\right\|$.  Methods for conversion to and analysis of the Kraus form are given in \cite{Havel2003} and \cite{Weinstein2004}.  The Kraus form for an ideal implementation of a unitary process would consist of a single Kraus operator which is the corresponding unitary operator describing the process.  Therefore, in an implementation where the errors are small, we expect the Kraus operator of largest magnitude to resemble the ideal unitary operator.  The numerically simulated superoperators expressed in the momentum basis, along with the largest magnitude Kraus operators, plotted in Fig. \ref{fig:sup_ops}, give a qualitative picture of the differences between a quantum process that leads to localization (in the ideal simulations) and a quantum process that causes some degree of delocalization (in the simulations which include errors).  
Off-diagonal elements in the unitary or Kraus operator cause transitions between momentum states; diagonal elements alter the magnitude and phase of each momentum state without causing transitions.  The qualitative result of the simulated noise is to reduce the bandedness (i.e. the relative magnitude of the diagonal and near-diagonal elements compared to the off-diagonal elements) of the operator, thus reducing the degree to which a state is localized under the map.  The simulated operators for 1 and 2 iterations are progressively less banded as more errors are included in the simulation, which further attests to the sensitivity of localization to experimental noise effects.  As explained in Sec. \ref{sec:num_sim}, Fig. \ref{fig:sup_ops} again shows qualitatively that rf inhomogeneity has a greater delocalizing effect than decoherence in this implementation of the quantum sawtooth map.

\section{Conclusions}
\label{sec:conclusions}

The quantum sawtooth map has been emulated on a three qubit liquid state nuclear magnetic resonance quantum information processor in the perturbative localization parameter regime of the map ($K=1.5$, $L=7$, $N=8$).  Observing the dynamic behavior of the width of the peak in the momentum probability distribution reveals behavior which is consistent with coherent quantum localization.  Due to incoherent noise, this localized peak is superimposed with a uniform background offset over the probability distribution which represents those parts of the ensemble which are not localized due to local unitary errors which vary over the ensemble.  Numerical simulations of the experiment reveal that the decoherent noise known to act on the system is relatively inconsequential in this implementation of the map, in terms of its effect on localization when compared with incoherence.  This study serves as a test of the capabilities of coherent control and serves to motivate the refinement of our implementation.  Specifically, we see that incoherence is the biggest challenge in implementing localization, a highly sensitive quantum coherence-dependent phenomenon.  

\section{Acknowledgements}
\label{sec:acknowl}

The authors would like to thank B. Levi, D. Poulin, M. Saraceno, G. Benenti and G. Casati for helpful discussions, N. Boulant for help with numerical simulations, L. Viola for bringing the feasibility of the experiment to our attention, and Y. Weinstein for help in preliminary studies.  This material is based upon work supported under a National Science Foundation Graduate Research Fellowship, in addition to support from ARO, ARDA, and LPS.   

\bibliography{Sawtooth_ps}
\end{document}